# Propagating Collective Spin-valley Modes in Twisted $WSe_2$

**Authors:** Richen Xiong[1], Yi Guo[1], Chenxin Qin[1], Taige Wang[2-4], Fanzhao Yin[1], Samuel L. Brantly[1], Youngjoon Choi[1], Junhang Qi[1], Jinfei Zhou[1], Zihan Zhang[1], Melike Erdi[5], Kenji Watanabe[6], Takashi Taniguchi[7], Shu Zhang[8], Seth Ariel Tongay[5], Andrea F. Young[1], Liang Fu[4], Chenhao Jin[1]*

**Affiliations:**

[1]*Department of Physics, University of California at Santa Barbara, Santa Barbara, CA, USA.*

[2]*Department of Physics, University of California at Berkeley, Berkeley, CA, USA.*

[3]*Department of Physics, Harvard University, Cambridge, MA, USA.*

[4]*Department of Physics, Massachusetts Institute of Technology, Cambridge, MA, USA.*

[5]*Materials Science and Engineering Program, School for Engineering of Matter, Transport, and Energy, Arizona State University, Tempe, AZ, USA.*

[6]*Research Center for Electronic and Optical Materials, National Institute for Materials Science, Tsukuba, Japan.*

[7]*Research Center for Materials Nanoarchitectonics, National Institute for Materials Science, Tsukuba, Japan.*

[8]*Collective Dynamics and Quantum Transport Unit, Okinawa Institute of Science and Technology Graduate University, Japan.*

* Corresponding author. Email: jinchenhao@ucsb.edu

**Abstract:**

The emergence of neutral collective modes is a hallmark of correlated quantum phases but is often challenging to probe experimentally. In two-dimensional flatband systems, charge responses have been intensively investigated yet neutral excitations remain largely unexplored. In particular, intervalley coherent state (IVC) features a neutral Goldstone mode due to spontaneously broken valley U(1) symmetry. While IVC state has been proposed as a unifying theme across graphene- and semiconductor-based systems, its defining feature – the neutral Goldstone mode – remains elusive in experiment. Here we investigate space-and-time-resolved transport of neutral modes in twisted $WSe_2$ moiré superlattices through a novel ultrafast imaging technique. We uncover two new propagating collective modes with very different velocities, which emerge near the van Hove singularity (VHS) in both intermediate- (3.5 to 4 degree) and large- (around 5 degree) angle twisted $WSe_2$. The fast-propagating mode has a large speed of about 3 km/s and is consistent with a Goldstone mode for an IVC state, while the slow-moving mode is likely a gapped amplitude mode. They can be understood as the spin-valley analogues of collective modes of a superfluid, whose propagation is imaged for the first time in a condensed matter system. Our study demonstrates a powerful new approach for probing charge-neutral modes in quantum materials and offers key insights into the interplay between charge and spin-valley physics in moiré superlattices.

Transport measurements are considered a gold standard for identifying states of matter by providing direct and quantitative information on low energy responses[1–5]. While electrical transport is widely used to investigate charge-carrying (quasi)particles[6–18], it remains an experimental challenge to measure transport of emergent charge-neutral modes, which lie at the heart of numerous quantum phenomena[19–21]. Recently, electrical transport revealed intriguing phases in van der Waals flatband systems, including (fractional) quantum anomalous Hall insulators[9–13] and superconductivity[15–18]. The neutral spin-valley degree of freedom is expected to play a crucial role in these phenomena as an essential ingredient of both topology and correlation[22–25]. Experimentally, however, pure spin-valley excitations and their interplay with correlated charge orders remain difficult to access when they are decoupled from charged quasiparticles.

A foremost example is intervalley-coherent (IVC) order, a correlated state spontaneously breaking valley U(1) symmetry. Originally proposed for moiré graphene, IVC state is now considered universal across flatband systems composed of graphene and transition metal dichalcogenides (TMD) and plays a central role in their phase diagrams[26–31]. Electrical and magnetometry measurements have provided signatures of IVC states through symmetry-broken gaps and field-tunability[26,32,33]. Direct microscopic evidence has so far relied on charge textures associated with IVC states and confined to scanning tunneling microscope (STM) studies of single-gated graphene systems[34,35]. In twisted TMD bilayers with spin-valley locking, however, IVC state features a coherent superposition between K and K' valleys of opposite spins[36]. This leads to a spin texture without appreciable charge density modulation, rendering conventional STM inapplicable.

The defining feature of IVC state is a linearly dispersing neutral Goldstone mode that accompanies spontaneous breaking of valley U(1) symmetry [36,37], which should lead to efficient transport of valley pseudospin[36]. Experimentally, nevertheless, this hallmark has not been observed in any system. Here we directly image transport of collective spin-valley modes in semiconducting moiré superlattices using an ultrafast imaging technique. Figure 1a illustrates the experimental configuration, where a pump pulse creates excitations in the device, whose evolution is detected by a delayed probe pulse (see Methods). The spatial profile of pump light can be engineered to either create homogeneous excitations to isolate dynamics without transport, or line-shaped excitations to investigate one-dimensional transport along the perpendicular direction (Fig. 1a). The probe light always covers the entire device and captures full images of the excitations without scanning. The charge and spin-valley excitations can be selectively probed through different detection channels (see Methods). This allows us to not only measure transport of charge-neutral modes but also in a space-and-time-resolved fashion. The latter becomes important when multiple modes coexist, as they would be averaged out in conventional transport measurements without space or time resolution.

## Emerging spin-valley modes in $tWSe_2$

Fig. 1b shows longitudinal resistance $R_{xx}$ of a 5-degree-twisted $WSe_2$ device D1 from electrical transport at our base temperature of 2.5 K (see Extended Data Fig. 1 for device images). Consistent with previous reports[17], no insulating state is observed at moiré filling $\nu$ = -1 (one hole per moiré unit cell), and the region near van Hove singularity (VHS) shows slightly larger resistance. We start by identifying spin-valley modes in the system. To this end, we employ a homogeneous wide-field pump and select the detection channel to be sensitive to out-of-plane spin-valley polarization $S_z$ (see Methods). Due to the optical selection rule of excitons in TMDs[38], spin-valley polarized excitons can be injected using circularly polarized light, which becomes pure spin-valley imbalance after electron-hole recombination[39]. Such a process is universal in semiconducting TMDs and does not involve correlated physics[39,40]. We therefore call pure spin-valley imbalance (of holes) an ordinary mode to underscore its single particle nature. As expected, a circular-polarized pump light creates this ordinary mode in the entire phase diagram of twisted $WSe_2$ (Fig. 1c). An unpolarized pump light, in contrast, does not generate net $S_z$, and therefore no signal is observed at zero magnetic fields (see Extended Data Fig. 2).

Interestingly, an additional signal emerges upon applying a small out-of-plane magnetic field (Fig. 1d). Unlike the ordinary mode, such signal does not depend on the pump light polarization and only appears around the VHS. We therefore call it exotic mode. The distinct pump polarization dependence between ordinary and exotic modes allows us to separate them by computing the symmetric and anti-symmetric responses to RCP and LCP (right- and left-circularly-polarized) pump; or equivalently, subtracting the response to unpolarized pump from that to RCP/LCP pump (see Methods and Extended Data Fig. 2). The obtained anti-symmetric responses ($(\Delta R/R)_{\text{asym}} = \left[(\Delta R/R)_{\text{RCP pump}} - (\Delta R/R)_{\text{LCP pump}}\right]/2$) at 0.5 T are almost identical to 0 T (Fig. 1e), indicating negligible effects of the small magnetic field on the ordinary mode over the entire phase diagram. The large amplitude of the exotic signal (symmetric responses, $(\Delta R/R)_{\text{sym}} = \left[(\Delta R/R)_{\text{RCP pump}} + (\Delta R/R)_{\text{LCP pump}}\right]/2$) at 0.5 T (Fig. 1d), comparable to the ordinary one, is therefore quite unusual. We further measure spin-valley relaxation dynamics of both the ordinary and exotic modes, as shown in Fig. 1f and 1g. Again, the dynamics of ordinary mode remain largely unchanged between 0 and 0.5 T (see 0 T results in Extended Data Fig. 2c), both featuring significant slower relaxation around VHS. The dynamics of the exotic mode are complicated, as will be discussed later.

## Unusual transport of exotic modes

We perform space-and-time resolved transport measurements in the spin-valley channel to investigate the nature of the exotic mode, so that we can rule out trivial enhancement around VHS due to, e.g. a larger magnetic susceptibility[17]. Fig. 2a shows six snapshots of space-time evolution of the ordinary mode at $n$ = -8.3×10$^{12}$ cm$^{-2}$, $E$ = 0.1 V/nm and $B_z$ = 0.5 T (star symbol in Fig. 1e, see Supplementary Video 1 and 2 for complete movie). At delay $\Delta t$ = 0, the initial excitations are homogeneous along the $y$-direction but confined to a diffraction-limited region in the $x$-direction. Afterwards, these excitations propagate out over time and become homogeneous over space at $\Delta t$ > 10 ns, limited by the sample boundary (black dashed line).

Transport of the exotic mode shows clearly distinct behaviors under identical experimental configurations (Fig. 2b). Notably, two exotic modes are observed, with drastically different propagation speed and opposite $S_z$. The fast exotic mode carries positive $S_z$ (aligned with the equilibrium magnetization) and reaches the sample boundary within $\Delta t$ = 1 ns, when the ordinary mode has only slightly expanded. In contrast, the slow exotic mode carries negative $S_z$ and reaches the sample boundary around $\Delta t$ = 35 ns, long after the ordinary mode. Their distinct speed allows them to be directly isolated in space-and-time resolved transport. In contrast, Fig. 1d only captures the total response of the two modes due to the lack of spatial resolution, which cannot identify or separate the two modes. The dynamics (Fig. 1f) is also a mix of the two modes and therefore complicated.

The unusual transport of the exotic modes, one faster and one slower than the ordinary mode, indicates their distinct nature. To allow quantitative analysis, Fig. 2c and 2d summarizes the time-dependent spatial profile along the $x$-direction of the ordinary and exotic modes, respectively. Transport of the ordinary mode is well captured by a diffusion-decay model (Fig. 2e and 2g, see Methods), from which we extract a diffusion constant of 15.2 cm$^2$/s and a lifetime of about 100 ns. The latter is consistent with dynamics measured without transport (Fig. 1g). We also attempt to fit transport of the exotic modes using a two-component diffusion decay model (see Methods), which yields a diffusion constant of 28.3 and 2.08 cm$^2$/s for the fast and slow modes, respectively. Nevertheless, the fit always deviates considerably from the experiment (Fig. 2f). In particular, the fit fails to capture the rapid propagation of the fast mode at the beginning. Fig. 2h compares spatial profile of the exotic modes to the best fit at representative delays within the first two nanoseconds. Focusing on the right half of the curves, the maximum of the fast mode (blue arrows) shows marked discrepancies from the fitting (orange arrows) and moves several times faster than the latter. Furthermore, their shapes are qualitatively different. While the fitted curve always shows a single local maximum, the experimental data shows two local maxima around $\Delta t$ = 1.4 ns. These observations suggest that transport of the fast mode likely possesses a non-diffusive component. One possible candidate is a ballistic component. Such a component manifests as a wave packet and leads to an additional local maximum in the spatial profile, which is consistent with the observation in Fig. 2h. However, we cannot quantitatively isolate transport of the non-diffusive component due to the limited sample size and the coexistence of multiple modes (see Methods for more discussions).

## Evolution of exotic modes

Fig. 3a and 3b shows the time-dependent spatial profile of the ordinary and exotic modes, respectively, at fixed hole doping $n = -8.3\times10^{12}$ cm$^{-2}$ and representative electric fields. Both the fast and slow exotic modes only appear adjacent to VHS, consistent with Fig. 1d. The close connection between the exotic modes and VHS is further confirmed by additional doping- and electric-field-dependent spin-valley transport measurements (see Extended Data Fig. 3), where both exotic modes disappear outside the VHS region. In contrast, the ordinary mode is observed at all electric fields, including the layer-polarized region (Fig. 3a and Extended Data Fig. 3). We have also performed temperature dependent spin-valley transport, as shown in Fig. 3c and 3d. Both fast and slow exotic modes show sensitive temperature dependence, disappearing at around 10 and 20 K, respectively. The amplitude of ordinary mode shows much weaker temperature dependence and remains largely unchanged up to 20 K. The lifetime of all modes decreases considerably with temperature, which is expected from stronger spin scattering at elevated temperatures[41].

To test universality of our observation, we performed similar measurements in two additional $tWSe_2$ devices D2 and D3 with smaller twist angles of 3.8 and 3.5 degrees. The longitudinal resistance of both devices (Fig. 4a and f) shows prominent insulating states at moiré filling $\nu$ = -1 and finite electric field, which become metallic at both small and large electric fields. Additional insulating states appear at $\nu$ = -1/3 under small electric field. These results are consistent with previous reports of $tWSe_2$ with twist angles of 3.5 to 4 degrees[15,16]. Spin-valley measurements of device D2 and D3 reproduce all observations from device D1. Exotic modes emerge around VHS with unpolarized pump excitation under a small $B_z$ (Fig. 4c and h). Spin-valley transport further reveals two exotic modes with opposite sign of $S_z$ and distinct transport speed (Fig. 4e and j, see Extended Data Fig. 6 for more results), one faster and one slower than the ordinary mode (Fig. 4d and i). The ordinary mode shows largely homogeneous amplitude throughout the phase diagram (Fig. 4b and g) with markedly longer lifetime at VHS (Extended Data Fig. 5). Interestingly, we do not observe strong signatures of the $\nu$ = -1 insulating state in any of the spin-valley measurements, despite their prominence in electrical transport. For example, the emergence of exotic modes and the enhanced lifetime of the ordinary mode both closely follow the shape of VHS and penetrate through $\nu$ = -1 without clear disruptions.

## Origin of the exotic modes

The distinct transport behaviors set the exotic modes apart from the ordinary mode and suggest a correlated origin. We further performed measurements with below-gap pump (Extended Data Fig. 10). The ordinary mode signal is no longer observed as the pump photon energy is insufficient to create interband particle-hole excitations. In contrast, the exotic modes signal remains similar, confirming their origins from low energy collective excitations. The fast mode is particularly striking as it shows fast propagation that cannot be fully captured by diffusion. A rough estimate from the first two nanoseconds' transport gives a velocity of about 3 km/s, assuming ballistic transport, or a diffusion constant of about 200 $cm^2/s$ if enforcing diffusive fit (see Extended Data Fig. 7). Such fast propagation requires the mode to have both a large group velocity and low energy and is therefore likely a Goldstone or near-gapless mode (see Methods for more discussions). Indeed, recent theoretical studies predicted intervalley coherent (IVC) states at VHS of twisted TMD moiré superlattices[27–30,36]. The phase of the coherent superposition $\theta$ is spontaneously fixed, which breaks the approximate intervalley $U(1)$ symmetry. This naturally gives rise to a Goldstone mode with group velocity in the order of 1 km/s (Ref. [31]), consistent with our experimental observation.

Another striking observation is the opposite sign of $S_z$ carried by the two exotic modes, which indicates that one of the modes does not couple linearly to $B_z$ in free energy but carries $S_z$ current. The IVC Goldstone mode does precisely that. We can make an analog to the superfluid, where $\theta$ plays the role as the superfluid phase, and $S_z$ resembles particle density[36]. Gradient of superfluid phase supports dissipationless density transport through supercurrent[42]. Similarly, the Goldstone mode of the IVC state here supports efficient $S_z$ transport through spin-valley supercurrent (see Methods for more discussions)[42].

Such a picture naturally captures all salient features of the experiments. Locally heating up a superfluid will partially convert the condensate into normal fluid, resulting in reduced superfluid density [43–45]. Such a deficit in the condensate will propagate out rapidly through the supercurrent, while the normal massive modes carrying opposite density diffuse slower. This reproduces our experimental observation, with the understanding that the particle number here is $S_z$. This intuition can be put more precisely by a minimum hydrodynamic theory, in which the phase and amplitude modes naturally carry opposite $S_z$ and exhibit ballistic and diffusive transport respectively (see Methods and Extended Data Fig. 7b). The gapped amplitude mode modulates the superfluid density and is also known as the Higgs mode[19]. Our space-and-time-resolved transport allows us to separate and capture both modes, which has so far only been achieved in synthetic systems such as ultracold atoms[44,45].

We further performed systematic magnetic field dependence to confirm origin of the exotic modes as collective excitations associated with an IVC state (Fig. 5). Signals from both fast and slow exotic modes (red and black symbols) show nonmonotonic magnetic field dependence and are suppressed at $B_z > 2T$, consistent with IVC collective modes as they should be suppressed when the system transitions into a valley-polarized state under large $B_z$. Other scenarios, such as paramagnetic responses of single particle excitations or phonons, should follow the system's equilibrium spin-valley polarization (blue symbols) and can be directly excluded (see Methods for detailed discussions).

## Discussion and outlook

Our results provide strong evidence for Goldstone modes and IVC states in both large (5-degree) and intermediate angle (3.5 to 4 degree) twisted $WSe_2$ moiré superlattices. In the 5-degree twisted sample, IVC signals trace the VHS and are strongest at intermediate electric fields (Fig. 1d). At smaller twist angles of 3.5 to 4 degree (Fig. 4, c and h), the strongest signal shifts to lower electric field. Superconductivity was previously reported in both cases[15–17] and is within or adjacent to the IVC states determined here. This is consistent with the potential connection between IVC states and superconductivity, as recently proposed by theoretical studies[29,46,47]. Interestingly, we do not observe clear effects of moiré filling $\nu = -1$ on spin-valley physics or IVC states, which suggests additional factors affecting superconductivity in smaller angle twisted $WSe_2$, such as other candidates of correlated insulators at $\nu = -1$ from moiré commensuration[48]. Since Goldstone mode is a generic property of IVC states, we cannot pinpoint the specific IVC types, including incommensurate Kekulé spiral (IKS) orders[49]. Quantitative comparison between experimental and theoretical spin-valley transport could further elucidate nature of the IVC states, which presents an exciting future direction that remains unexplored so far.

More generally, our study sets a powerful approach for investigating charge neutral modes through space-and-time-resolved transport. It allows us to directly resolve the emergence of multiple modes and monitor their individual transport, such as the phase and amplitude modes of a spin-valley superfluid here. This is highly challenging in conventional transport, where the measurement would be short-circuited by the most conducting mode. Our approach opens up opportunities for studying many intriguing phenomena, such as excitonic insulator, quantum spin liquid and fractional quantum anomalous Hall insulators, where multiple low energy modes coexist, among which many are charge neutral[21,50–53].

## Reference:

1. Pekola, J. P. & Karimi, B. Colloquium : Quantum heat transport in condensed matter systems. *Rev. Mod. Phys.* **93**, 041001 (2021).

2. Ma, Q., Krishna Kumar, R., Xu, S. Y., Koppens, F. H. L. & Song, J. C. W. Photocurrent as a multiphysics diagnostic of quantum materials. *Nature Reviews Physics* **5**, 170–184 (2023).

3. Varnavides, G., Yacoby, A., Felser, C. & Narang, P. Charge transport and hydrodynamics in materials. *Nat. Rev. Mater.* **8**, 726–741 (2023).

4. Han, W., Maekawa, S. & Xie, X.-C. Spin current as a probe of quantum materials. *Nat. Mater.* **19**, 139–152 (2020).

5. Yuan, L. *et al.* Twist-angle-dependent interlayer exciton diffusion in WS2–WSe2 heterobilayers.

*Nat. Mater.* **19**, 617–623 (2020).

6. Cao, Y. *et al.* Unconventional superconductivity in magic-angle graphene superlattices. *Nature* **556**, 43–50 (2018).

7. Hao, Z. *et al.* Electric field–tunable superconductivity in alternating-twist magic-angle trilayer graphene. *Science.* **371**, 1133–1138 (2021).

8. Lu, X. *et al.* Superconductors, orbital magnets and correlated states in magic-angle bilayer graphene. *Nature* **574**, 653–657 (2019).

9. Cai, J. *et al.* Signatures of fractional quantum anomalous Hall states in twisted MoTe2. *Nature* **622**, 63–68 (2023).

10. Lu, Z. *et al.* Fractional quantum anomalous Hall effect in multilayer graphene. *Nature* **626**, 759–764 (2024).

11. Park, H. *et al.* Observation of fractionally quantized anomalous Hall effect. *Nature* **622**, 74–79 (2023).

12. Zeng, Y. *et al.* Thermodynamic evidence of fractional Chern insulator in moiré MoTe2. *Nature* **622**, 69–73 (2023).

13. Xu, F. *et al.* Observation of Integer and Fractional Quantum Anomalous Hall Effects in Twisted Bilayer MoTe2. *Phys. Rev. X* **13**, 031037 (2023).

14. Wang, Y. *et al.* Hidden states and dynamics of fractional fillings in twisted MoTe2 bilayers. *Nature* **641**, 1149–1155 (2025).

15. Wang, L. *et al.* Correlated electronic phases in twisted bilayer transition metal dichalcogenides. *Nat. Mater.* **19**, 861–866 (2020).

16. Xia, Y. *et al.* Superconductivity in twisted bilayer WSe2. *Nature* **637**, 833–838 (2025).

17. Guo, Y. *et al.* Superconductivity in 5.0° twisted bilayer WSe2. *Nature* **637**, 839–845 (2025).

18. Xu, F. *et al.* Signatures of unconventional superconductivity near reentrant and fractional quantum anomalous Hall insulators. arXiv: 2504.06972 (2025).

19. Pekker, D. & Varma, C. M. Amplitude/Higgs Modes in Condensed Matter Physics. *Annu. Rev. Condens. Matter Phys.* **6**, 269–297 (2015).

20. Savary, L. & Balents, L. Quantum spin liquids: a review. *Reports on Progress in Physics* **80**, 016502 (2017).

21. Wu, S. *et al.* Charge-neutral electronic excitations in quantum insulators. *Nature* **635**, 301–310 (2024).

22. Wu, F., Lovorn, T., Tutuc, E., Martin, I. & Macdonald, A. H. Topological Insulators in Twisted Transition Metal Dichalcogenide Homobilayers. *Phys. Rev. Lett.* **122**, 086402 (2019).

23. Devakul, T., Crépel, V., Zhang, Y. & Fu, L. Magic in twisted transition metal dichalcogenide bilayers. *Nat. Commun.* **12**, 6730 (2021).

24. Jia, Y. *et al.* Moiré fractional Chern insulators. I. First-principles calculations and continuum models of twisted bilayer MoTe2. *Phys. Rev. B* **109**, 205121 (2024).

25. Zhang, X.-W. *et al.* Polarization-driven band topology evolution in twisted MoTe2 and WSe2. *Nat. Commun.* **15**, 4223 (2024).

26. Bultinck, N. *et al.* Ground State and Hidden Symmetry of Magic-Angle Graphene at Even Integer Filling. *Phys. Rev. X* **10**, 031034 (2020).

27. Hsu, Y.-T., Wu, F. & Das Sarma, S. Spin-valley locked instabilities in moiré transition metal dichalcogenides with conventional and higher-order Van Hove singularities. *Phys. Rev. B* **104**, 195134 (2021).

28. Zang, J., Wang, J., Cano, J. & Millis, A. J. Hartree-Fock study of the moiré Hubbard model for twisted bilayer transition metal dichalcogenides. *Phys. Rev. B* **104**, 075150 (2021).

29. Fischer, A. *et al.* Theory of intervalley-coherent AFM order and topological superconductivity in tWSe2. arXiv: 2412.14296 (2024).

30. Peng, L. *et al.* Magnetism in Twisted Bilayer WSe2. arXiv:2503.09689 (2025).

31. Wang, T., Devakul, T., Zaletel, M. P. & Fu, L. Diverse magnetic orders and quantum anomalous Hall effect in twisted bilayer MoTe2 and WSe2. arXiv:2306.02501 (2023).

32. Ghiotto, A. *et al.* Stoner instabilities and Ising excitonic states in twisted transition metal dichalcogenides. arXiv:2405.17316 doi:10.48550/arXiv.2405.17316.

33. Arp, T. *et al.* Intervalley coherence and intrinsic spin–orbit coupling in rhombohedral trilayer graphene. *Nat. Phys.* **20**, 1413–1420 (2024).

34. Nuckolls, K. P. *et al.* Quantum textures of the many-body wavefunctions in magic-angle graphene. *Nature* **620**, 525–532 (2023).

35. Kim, H. *et al.* Imaging inter-valley coherent order in magic-angle twisted trilayer graphene. *Nature* **623**, 942–948 (2023).

36. Bi, Z. & Fu, L. Excitonic density wave and spin-valley superfluid in bilayer transition metal dichalcogenide. *Nat. Commun.* **12**, 642 (2021).

37. Kumar, A., Xie, M. & MacDonald, A. H. Lattice collective modes from a continuum model of magic-angle twisted bilayer graphene. *Phys. Rev. B* **104**, 035119 (2021).

38. Xu, X., Yao, W., Xiao, D. & Heinz, T. F. Spin and pseudospins in layered transition metal dichalcogenides. *Nat. Phys.* **10**, 343–350 (2014).

39. Jin, C. *et al.* Imaging of pure spin-valley diffusion current in WS2 -WSe2 heterostructures. *Science.* **360**, 893–896 (2018).

40. Yang, L. *et al.* Long-lived nanosecond spin relaxation and spin coherence of electrons in monolayer MoS2 and WS2. *Nat. Phys.* **11**, 830–834 (2015).

41. Jin, C. *et al.* Ultrafast dynamics in van der Waals heterostructures. *Nat. Nanotechnol.* **13**, 994–1003 (2018).

42. Sonin, E. B. Superfluid spin transport in magnetically ordered solids. *Low Temperature Physics* **46**, 436–447 (2020).

43. Andrews, M. R. *et al.* Propagation of Sound in a Bose-Einstein Condensate. *Phys. Rev. Lett.* **79**, 553–556 (1997).

44. Meppelink, R., Koller, S. B. & van der Straten, P. Sound propagation in a Bose-Einstein condensate at finite temperatures. *Phys. Rev. A.* **80**, 043605 (2009).

45. Sidorenkov, L. A. *et al.* Second sound and the superfluid fraction in a Fermi gas with resonant interactions. *Nature* **498**, 78–81 (2013).

46. Schrade, C. & Fu, L. Nematic, chiral, and topological superconductivity in twisted transition metal dichalcogenides. *Phys. Rev. B* **110**, 035143 (2024).

47. Zhu, J., Chou, Y.-Z., Xie, M. & Das Sarma, S. Superconductivity in twisted transition metal dichalcogenide homobilayers. *Phys. Rev. B* **111**, L060501 (2025).

48. Kim, S., Mendez-Valderrama, J. F., Wang, X. & Chowdhury, D. Theory of correlated insulators and superconductor at $\nu = 1$ in twisted WSe2. *Nat. Commun.* **16**, 1701 (2025).

49. Kwan, Y. H. *et al.* Kekulé Spiral Order at All Nonzero Integer Fillings in Twisted Bilayer Graphene. *Phys. Rev. X* **11**, 041063 (2021).

50. Paul, N., Abouelkomsan, A., Reddy, A. & Fu, L. Shining light on collective modes in moir\'e fractional Chern insulators. arXiv:2502.17569 (2025).

51. Kousa, B. M., Morales-Durán, N., Wolf, T. M. R., Khalaf, E. & MacDonald, A. H. Theory of magnetoroton bands in moir\'e materials. arXiv: 2502.17574 (2025).

52. Qiu, W.-X. & Wu, F. Topological magnons and domain walls in twisted bilayer MoTe2. arXiv:2502.11010 (2025).

53. Zhou, W.-T., Dong, Z.-Y., Gu, Z.-L. & Li, J.-X. Itinerant topological magnons and spin excitons in twisted transition metal dichalcogenides: Mapping electron topology to spin counterpart.

arXiv:2502.10991 (2025).

54. Wang, L. *et al.* One-dimensional electrical contact to a two-dimensional material. *Science.* **342**, 614–617 (2013).

55. Xiong, R. *et al.* Correlated insulator of excitons in WSe2/WS2 moiré superlattices. *Science.* **380**, 860–864 (2023).

56. Song, W. *et al.* High-Resolution Van der Waals Stencil Lithography for 2D Transistors. *Small* **17**, (2021).

57. Jin, C. *et al.* Imaging and control of critical fluctuations in two-dimensional magnets. *Nat. Mater.* **19**, 1290–1294 (2020).

58. Jin, C. *et al.* Stripe phases in WSe2/WS2 moiré superlattices. *Nat. Mater.* **20**, 940–944 (2021).

59. Regan, E. C. *et al.* Spin transport of a doped Mott insulator in moiré heterostructures. *Nat. Commun.* **15**, 10252 (2024).

60. Bauer, G. E. W., Saitoh, E. & van Wees, B. J. Spin caloritronics. *Nat. Mater.* **11**, 391–399 (2012).

61. Zhang, W. *et al.* Determination of the Pt spin diffusion length by spin-pumping and spin Hall effect. *Appl. Phys. Lett.* **103**, (2013).

62. Ko, K.-H. & Choi, G.-M. Optical method of determining the spin diffusion length of ferromagnetic metals. *J. Magn. Magn. Mater.* **510**, 166945 (2020).

63. Yu, H. *et al.* High propagating velocity of spin waves and temperature dependent damping in a CoFeB thin film. *Appl. Phys. Lett.* **100**, 262412 (2012).

64. Takei, S. & Tserkovnyak, Y. Superfluid Spin Transport Through Easy-Plane Ferromagnetic Insulators. *Phys. Rev. Lett.* **112**, 227201 (2014).

65. Takei, S., Halperin, B. I., Yacoby, A. & Tserkovnyak, Y. Superfluid spin transport through antiferromagnetic insulators. *Phys. Rev. B* **90**, 094408 (2014).

66. Feldman, D. E. & Halperin, B. I. Fractional charge and fractional statistics in the quantum Hall effects. *Reports on Progress in Physics* **84**, 076501 (2021).

67. Kousa, B. M., Morales-Durán, N., Wolf, T. M. R., Khalaf, E. & MacDonald, A. H. Theory of magnetoroton bands in moiré materials. arXiv:2502.17574 (2025).

68. Wolf, T. M. R., Xie, T., Jin, C. & MacDonald, A. H. Exciton-based sensing of remote electron correlations in 2D heterostructures. arXiv:2510.21522(2025).

69. Xie, T. *et al.* Optical imaging of flavor order in flat band graphene. *Nat. Commun.* **16**, 5555 (2025).

**Figure 1**

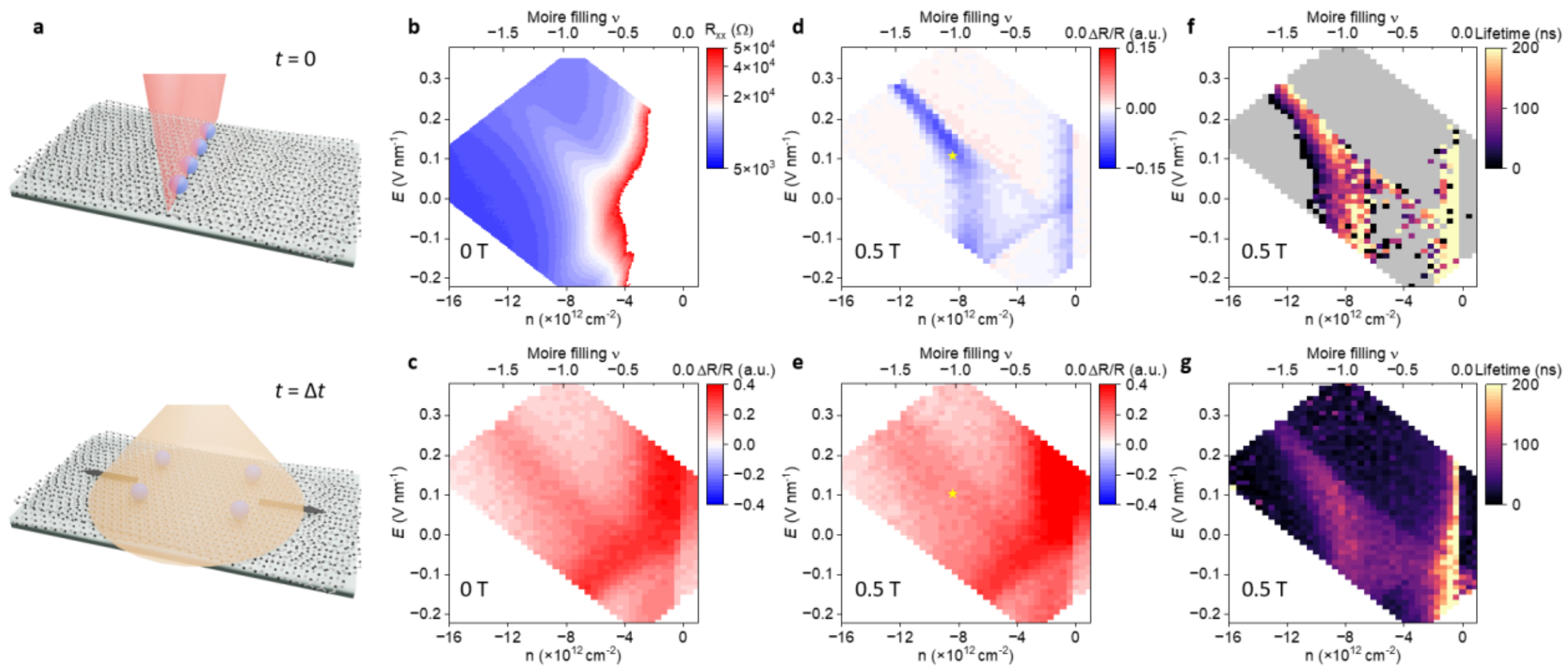

**Figure 1. Ultrafast imaging of spin-valley modes in 5° twisted $WSe_2$. a,** Illustration of ultrafast wide-field imaging. Top: an ultrafast pump pulse of 1.88 eV creates excitations either homogeneously for studying dynamics (not shown) or in a line-shape for studying 1D transport (shown here). Bottom: after delay $\Delta t$, a wide-field probe pulse of 1.67 eV monitors excitations with both space and time resolution. **b**, Longitudinal resistance $R_{xx}$ (two-probe) as a function of doping $n$ and electric field $E$ at zero magnetic field. The negative sign in $n$ represents hole doping. Resistance shows local enhancements near the van Hove singularity (VHS). **c,** Out-of-plane spin-valley polarization $S_z$ injected by an RCP pump light at $\Delta t$ = 1 ns and zero magnetic field. The signal ΔR/R is proportional to pump-induced magnetic circular dichroism and therefore $S_z$ (see Methods) **d**,**e**, Symmetric (**d**) and anti-symmetric (**e**) signals between RCP and LCP pump lights at $\Delta t$ = 1 ns and out-of-plane magnetic field $B_z$ = 0.5 T, which correspond to the exotic and ordinary modes, respectively. The ordinary mode (**e**) shows largely homogeneous responses over the phase diagram and negligible changes under magnetic field. In contrast, the exotic mode (**d**) emerges only around VHS and is absent without $B_z$. **f,g,** Spin-valley relaxation lifetime of exotic (**f**) and ordinary (**g**) modes across the ($n$, $E$) phase space at $B_z$ = 0.5 T. Both lifetimes show marked enhancement around VHS. All measurements are performed at the base temperature of 2.5 K in device D1.

**Figure 2**

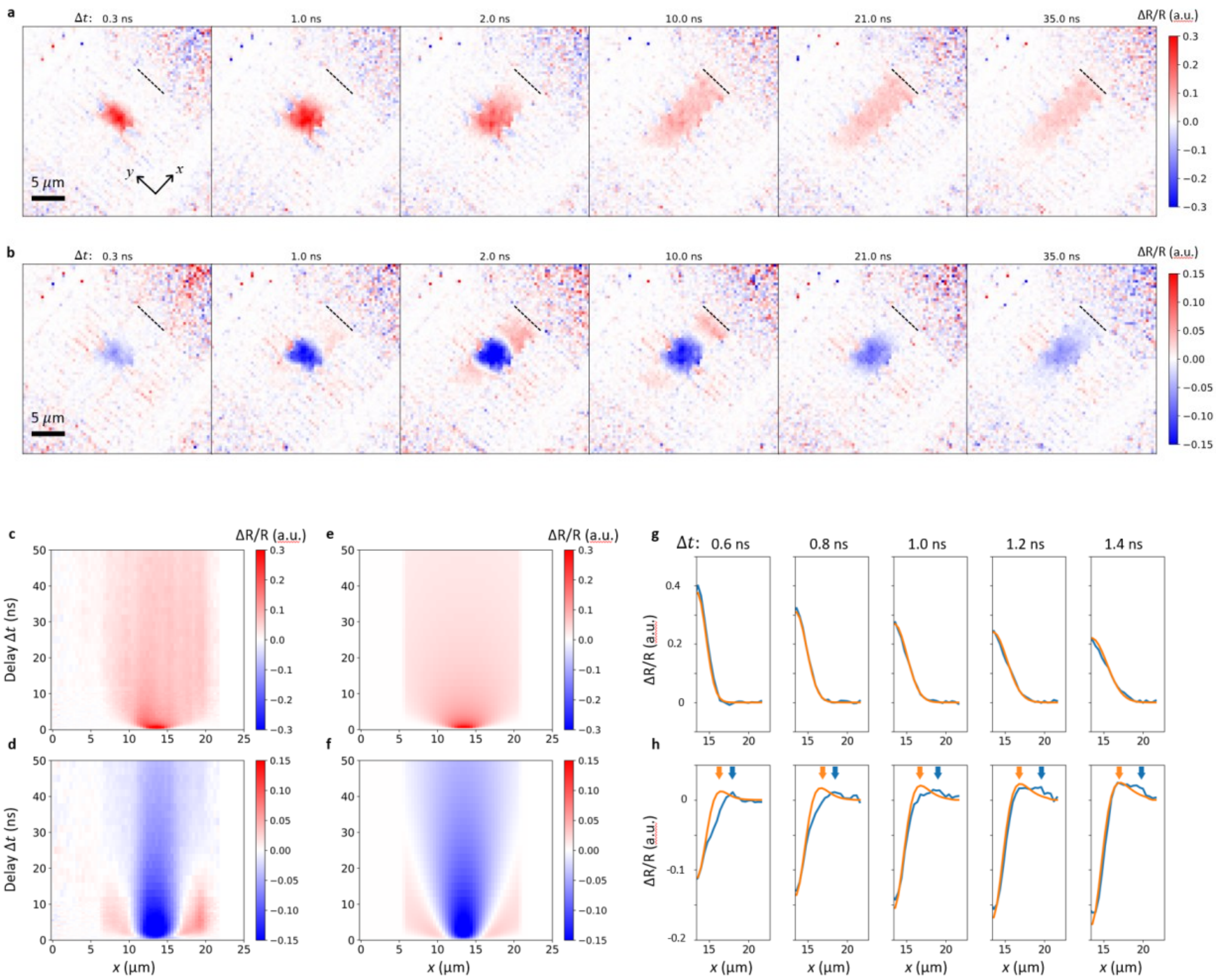

**Figure 2. Space-and-time resolved transport of spin-valley modes. a,b**, Space-time evolution of the ordinary (**a**) and exotic modes (**b**) at $n = -8.3\times10^{12}$ cm$^{-2}$, $E = 0.1$ V nm$^{-1}$, $B_z = 0.5$ T and representative delays. The excitations are launched by a line-shaped pump light in the center, which subsequently propagates along the $x$-direction until reaching the sample boundary (black dashed line, see Extended Data Fig. 1e-g for comparison with sample image). Two exotic modes are observed with opposite $S_z$ and very different velocities. The asymmetric propagation along positive and negative $x$ direction is due to spatial inhomogeneity within the device. **c-f**, Delay-dependent spatial profile of the ordinary (**c**) and exotic (**d**) modes along $x$ direction and the corresponding best fits using one- (**e**) and two-component (**f**) diffusion-decay model, respectively. **g,h,** Comparison between experimental and fitted spatial profile in the first 1.4 ns for ordinary (**g**) and exotic (**h**) modes. The ordinary mode is well-described by the diffusion-decay mode (**g**). In contrast, the fit fails to capture the rapid propagation of the fast mode (**h**), resulting in marked discrepancies between maxima in the fitted (orange arrows) and experimental (blue arrows) spatial profiles.

**Figure 3**

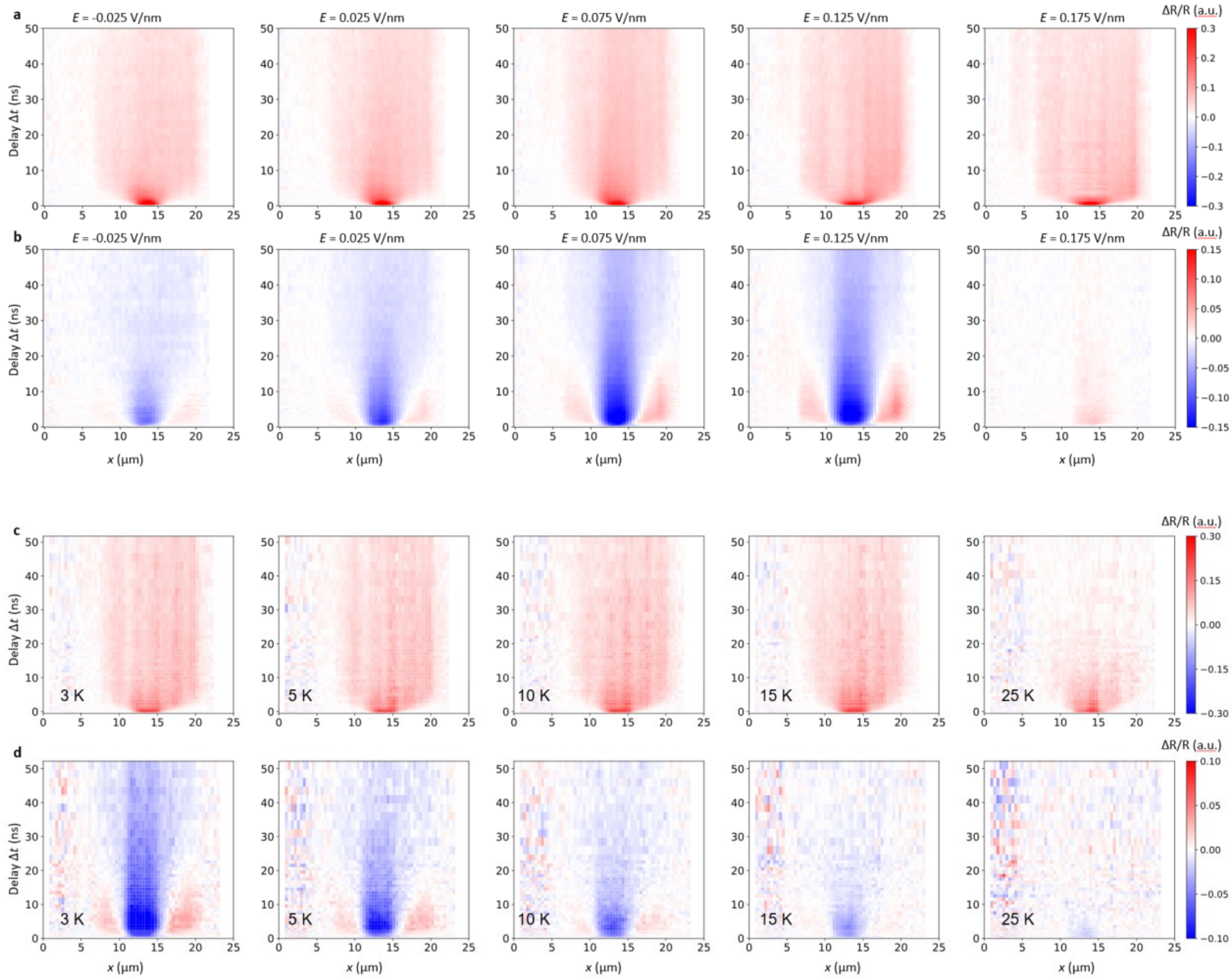

**Figure 3. Electric field and temperature dependence. a,b,** Delay-dependent spatial profiles at $n = -8.3\times10^{12}$ cm$^{-2}$, $B_z = 0.5$ T and representative electric field for the ordinary (**a**) and exotic (**b**) modes. The exotic modes emerge only around VHS, whereas the ordinary mode exists across all electric fields. **c,d,** Delay-dependent spatial profiles at $n = -8.3\times10^{12}$ cm$^{-2}$, $E = 0.1$ V nm$^{-1}$, $B_z = 0.5$ T and different temperatures. The ordinary mode's amplitude remains largely unchanged up to 25 K (**c**), while the fast and slow exotic modes (**d**) disappear above T ≈ 10 K and T ≈ 20 K, respectively.

**Figure 4**

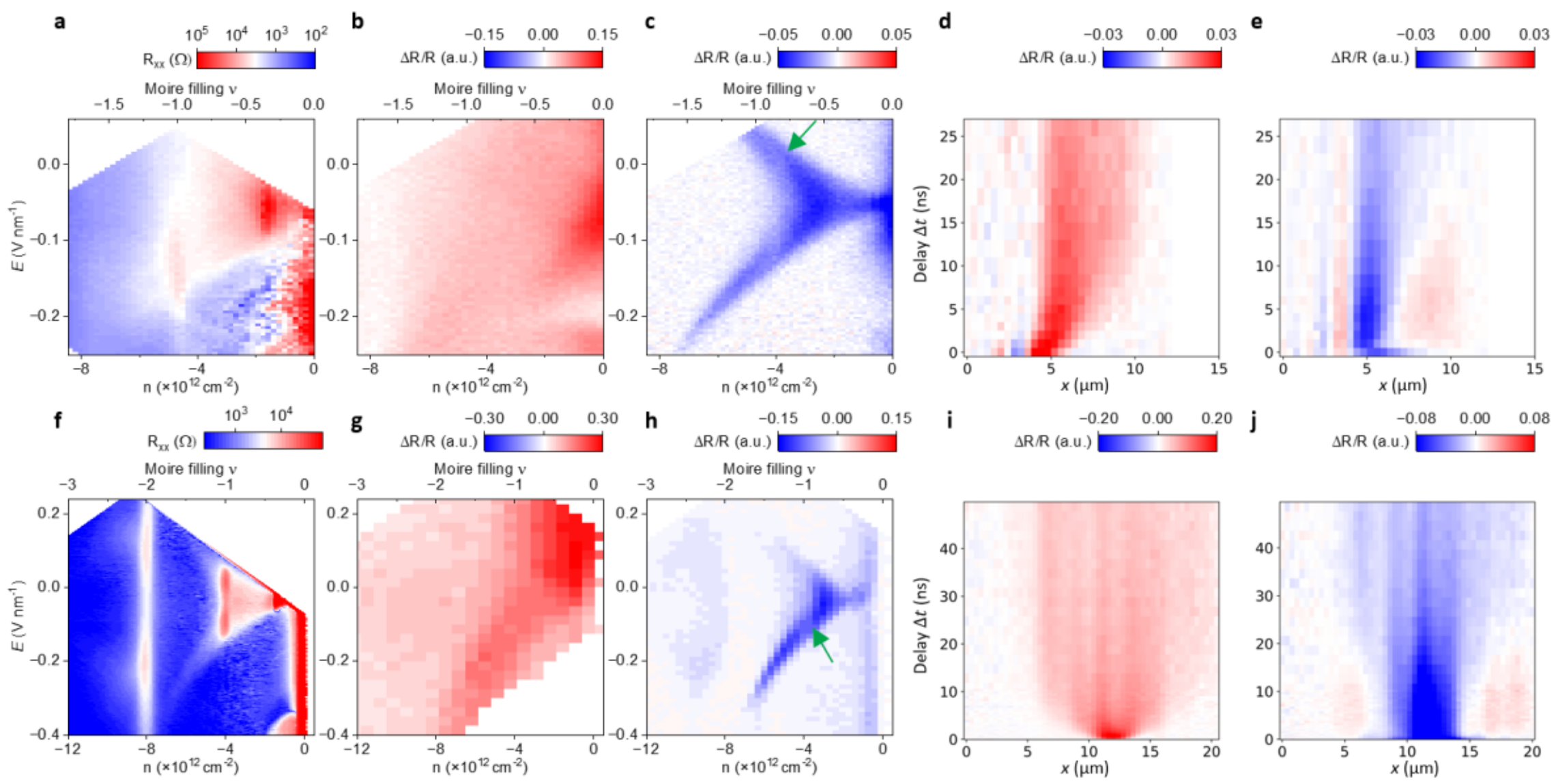

**Figure 4. Results from additional devices D2 (3.8°) and D3 (3.5°). a,** Longitudinal resistance $R_{xx}$ (four-probe) as a function of $n$ and $E$ at zero magnetic field for device D2. Correlated insulating peaks are observed at moiré fillings $\nu$ = -1 and $\nu$ = -1/3. **b, c,** Amplitude of ordinary (**b**) and exotic (**c**) modes at $\Delta t$ = 1 ns and $B_z$ = 0.5 T. **d,e,** Delay-time-dependent spatial profile of the ordinary mode (**d**) and exotic modes (**e**) at $n$ and $E$ indicated by the green arrow in (**c**). **f-j**, same as **a-e** for device D3. Correlated insulating peaks are observed at moiré fillings $\nu$ = -2 and $\nu$ = -1. In both devices, two exotic modes emerge around VHS with opposite $S_z$ and distinct velocities, consistent with the results from device D1. Interestingly, $\nu$= -1 does not show strong features in spin-valley measurements despite prominent insulating peaks in charge transport.

**Figure 5**

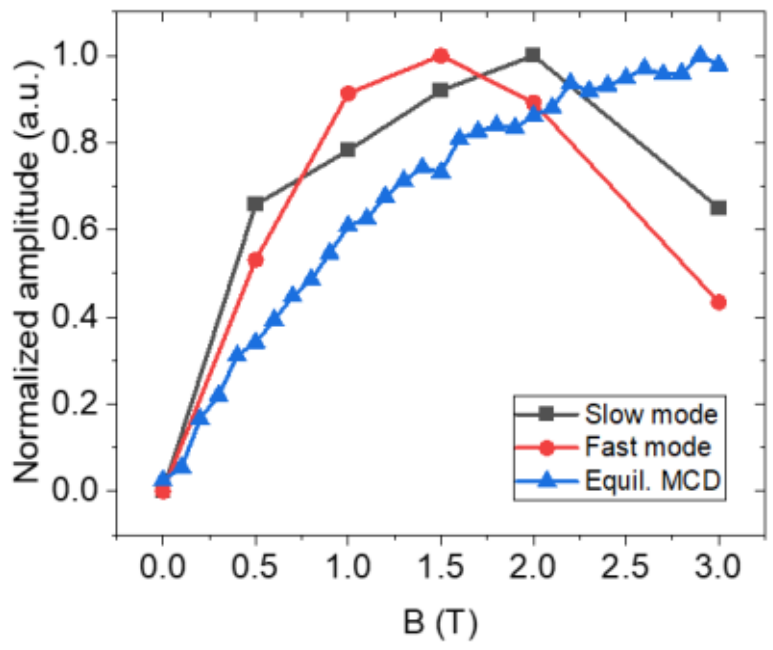

**Figure 5. Magnetic field dependence.** Magnetic field dependent amplitude of the slow and fast exotic modes compared with the sample's equilibrium MCD at $n = -8.3\times10^{12}$ cm$^{-2}$ and $E = 0.1$ V nm$^{-1}$. The curves are normalized by maximum (or minimum) for comparison. While equilibrium MCD increases monotonically with magnetic field, the exotics modes show a non-monotonic magnetic field dependence and are suppressed at large magnetic field.

**Methods:**

**Sample preparation:** The dual-gated twisted $WSe_2$ devices were made by layer-by-layer dry transfer method[54] of van der Waals (vdW) materials as described in previous studies[55]. The $WSe_2$ flake was cut into two halves using an STM tip (P-50 PtIr) on the transfer station. The flakes were picked up by a polycarbonate (PC) thin film on a polydimethylsiloxane stamp in the following sequence: hBN, top-gate graphite, hBN, first half of the $WSe_2$, second half of the $WSe_2$ with a controlled twist. The stack was then released to a bottom gate stack (hBN and bottom-gate graphite) with pre-patterned platinum electrodes. To achieve low-resistance electrical contacts at cryogenic temperature and preserve the surface cleanliness of vdW materials, we fabricate the bottom gate with platinum electrodes by flipping[35] and employing a PMMA stencil mask[56] for metal deposition, as detailed below.

First, we prepare a bottom gate stack assisted by a gold-coated PDMS block (Extended Data Fig. 1d, upper panel). We pick up graphite and hBN sequentially with PC film. PC film supporting the stack is then peeled off and placed, with the stack facing downward, onto a PDMS block coated with a thin layer of Ti/Au (3/12 nm). The PC film is then dissolved using N-methyl-2-pyrrolidone (NMP), leaving the stack supported on the gold surface. During this process, the stack's top surface remains protected from contamination by the evaporated metal layer. The structure is subsequently transferred onto an oxide substrate (Si with 90 nm $SiO_2$).

Next, a PMMA film with designed electrode pattern is fabricated on a bare silicon wafer pre-treated with hexamethyldisilazane (HMDS), using standard e-beam lithography (Extended Data Fig. 1d, middle panel). The PMMA stencil is picked up with a PMDS stamp and released onto the bottom gate stack, followed by standard e-beam evaporation of Cr/Pt (1/5nm). After the metal deposition, the whole PMMA film is peeled off from the substrate, thereby avoiding polymer residues associated with conventional solvent-based lift-off. The remaining layers of the vdW heterostructure are then placed onto the prepared bottom gate. After encapsulation, the local contact gate and contacts to the prepattern Pt electrodes are fabricated using standard e-beam lithography and metal deposition. Using this method, we are able to achieve contact resistance of $< 10\ k\Omega$ at cryogenic temperature when applying a large contact gate voltage.

**Ultrafast pump probe imaging:** All measurements were performed in a closed-cycle cryostat (Quantum Design, Opticool) at a base temperature of 2.5 K unless specified. Femtosecond pulses (1030 nm, 600 kHz, about 200 fs) were produced by seeding a regenerative amplifier with a mode-locked oscillator (Light Conversion). These pulses were subsequently guided into an optical parametric amplifier (Light Conversion) which produces pulse trains of 660 nm light as pump light. The probe light was produced from a fiber-based supercontinuum laser (SC, YSL Photonics). The SC (600 kHz, 100 ps) was triggered by an electronic pulse sourced from the mode-locked oscillator. Relative delay between pump and probe pulses was achieved by feeding the electronic pulse through digital delay generator (Hewlett-Packard 81110a) prior to triggering the SC. The SC pulses were further wavelength selected by home-built double monochromators to have a center wavelength around 740 nm with 0.2 nm full width at half max. The energy of the probe light is fine-tuned to maximize the signal for each device based on the $WSe_2$ 1s exciton resonance. Both pump and probe beams were independently expanded using imaging lenses then combined at a beamsplitter before impinging onto the sample. A cylindrical lens was used to shape the pump into a line profile at the sample plane. The reflected probe light was isolated from the pump light via a 700 nm long pass filter and collected on an electron-multiplying CCD camera (Princeton Instruments ProEM). In order to obtain the pump-induced change in the probe reflection, the pump was modulated by an optical chopper at 72 Hz, which was synchronized with the CCD camera. Comparison of images taken between pump "on" and "off" states allowed us to isolate relative reflection change as $\Delta R/R = (R_{pump\text{-}on} - R_{pump\text{-}off})/R_{pump\text{-}off}$. The typical power intensity we use is 0.5 nW/$\mu$m$^2$ for probe light and 10 nW/$\mu$m$^2$ for pump light. All optical measurements are performed in device D1 unless otherwise specified.

**Selective probe of spin-valley excitations**: The charge and spin-valley excitations can be selectively probed through different polarization configurations of the probe beam, as described in previous studies[57,58]. In brief, a pair of Glan–Thompson polarizers were used in combination to define and analyze the probe polarization, together with a broadband half-wave plate (HWP) and quarter-wave plate (QWP) positioned before the analyzer. Prior to interaction with the sample, the probe beam was linearly polarized and can be expressed as a coherent superposition of left- and right-circularly polarized (LCP and RCP) components with equal amplitude and phase[57]:

$$E_{r0}\begin{pmatrix}1\\0\end{pmatrix} = \frac{E_{r0}}{2}\begin{pmatrix}1\\i\end{pmatrix} + \frac{E_{r0}}{2}\begin{pmatrix}1\\-i\end{pmatrix} \qquad (1)$$

To probe charge excitations, we employed a "parallel" detection configuration, where the transmission axes of both polarizers were aligned. This configuration detects the total reflected intensity of LCP and RCP components, allowing optical reflection to capture changes in the dielectric function due to charge (i.e. population) dynamics.

In contrast, to probe spin-valley excitations, we used a "near-cross" detection configuration, where the polarizers' transmission axes are close to orthogonal. A finite $S_z$ in the system induces magnetic circular dichroism (MCD), i.e., imbalance between LCP and RCP reflection. This leads to an electric field component perpendicular to the incident light, which can be sensitively picked up in the near-cross detection configuration. Quantitatively, the MCD-induced relative reflection change is given by[57]:

$$\frac{\Delta R}{R} \approx \frac{2}{\phi}\alpha \tag{2}$$

Here α denotes the MCD amplitude, and $\phi$ is the angle between the two polarizers ($\phi = 0$ corresponds to orthogonal). The MCD signal can be enhanced by orders of magnitude by choosing a small $\phi$. In our measurements $\phi$ = 1°. Eq. (2) can also be used to identify the origin of the signal, as $\Delta R/R$ from spin-valley excitations would change signs when reversing the sign of $\phi$, while signals from charge (population) excitations are independent of $\phi$. We thereby verify that the observed ordinary and exotic signals are all from spin-valley excitations (see Extended Data Fig. 8).

The distinction between charge and spin-valley excitation is further confirmed by their different dynamics. At a representative $n = -8.3\times10^{12}$ cm$^{-2}$, $E = 0.1$ V nm$^{-1}$, $B_z = 0$ T and under RCP pump, the charge (population) lifetime (Extended Data Fig. 9a, black curve) is only a few nanoseconds. In contrast, the lifetime of spin-valley polarization (red curve) is much longer, at least 80 ns. Similar behaviors are observed at $B_z = 0.5$ T, under unpolarized pump (Extended Data Fig. 9b). These results directly confirm that all three modes discussed in the main text, one "ordinary" mode and two "exotic" modes, are pure spin-valley modes decoupled from charge (population) excitations.

**Below-gap pump measurements:** To distinguish low-energy collective spin excitations from single-particle excitations, we performed additional measurements using a below-gap (900 nm) pump while keeping the probe configuration unchanged. Extended Data Fig. 10, a and b show delay-dependent spatial profile of ordinary and exotic modes in spin-transport, respectively. With below-gap pumps, ordinary mode is no longer observed since pump photon does not have sufficient energy to create interband transitions or inject single-particle excitations. In contrast, both fast and slow exotic modes can still be excited and show transport behaviors consistent with that under above-gap pump. The contrasting behaviors between the ordinary and exotic modes indicate their distinct nature: while the ordinary mode is single particle excitations injected from interband high energy particle-hole excitations, the exotic modes are low energy collective excitations created by heating and/or intraband transitions.

**Non-diffusive transport of the fast mode:** To quantify the spatial-temporal evolution of spin-valley modes, we fit the spatial profiles along the transport direction $x$ using a diffusive-delay model[39,59]:

$$\Delta p(x,t) = \frac{p_0}{\sqrt{\pi(\sigma_0^2 + 4Dt)}} e^{-\frac{x^2}{\sigma_0^2+4Dt}} e^{-\frac{t}{\tau}} \quad (3)$$

where $p_0$ is the total number of pump-induced spin-imbalance, $\sigma_0$ is half width of the pump beam ($\sigma_0 \approx 1$ μm), $D$ is the spin diffusion constant, and $\tau$ is the spin lifetime. The ordinary mode can be fit well with the diffuse-delay model (Fig. 2g), from which we extract both diffusion constant and lifetime.

For the exotic modes, we employ a two-component version of the diffusion-decay model to account for the coexistence of two modes with opposite sign (opposite $S_z$) and different velocities. To make the fitting more stable, we first fit the slow-moving exotic mode using a single-component model and use the extracted parameters as initial guesses for the full two-component fit. This approach captures the evolution of the slow mode. However, the fast-propagating mode is not well captured by this model and travels several times faster than the best fit (Fig. 2h). We have also attempted to fit the fast mode with an unphysically large diffusion constant (Extended Data Fig. 7). However, the model again fails to capture the spatial-time evolution of the exotic modes.

Our results therefore indicate (at least partially) non-diffusive transport of the fast mode, which, together with the large velocity, suggests a Goldstone or near-gapless mode nature. The observations here require the mode to have both weak scattering and a large group velocity. These two conditions are typically incompatible for a gapped magnon mode in metals, since large group velocity necessitates large momentum and high energy, which leads to strong scattering. Indeed, gapped magnons in metallic thin films typically show diffusive transport with diffusion length of a few nm[60–63]. In contrast, a Goldstone mode naturally satisfies these requirements as the linear dispersion leads to a large group velocity regardless of momentum and the gapless nature allows arbitrarily soft modes. In the limit of no scattering, the Goldstone mode would move ballistically as a wave packet, maintaining a relatively sharp front with minimal broadening, which gives rise an additional local maximum in the spatial profile (Fig. 2h).

On the other hand, it is challenging to quantitatively analyze transport of the non-diffusive component. First, its isolation is complicated by the coexistence of two spatially overlapping modes with opposite spin-valley polarization ($S_z$). Consequently, we cannot reliably determine the width or shape of the two modes. Second, the propagation of collective modes is sensitive to disorders and inhomogeneities such as defects, twist-angle variations and sample boundaries. As a result, the transport of a Goldstone mode may not be completely ballistic and is likely more complicated. For example, it may show a mix of diffusive and ballistic transport.

**Analogy between intervalley coherent (IVC) states and superfluid:** An intervalley coherent state (IVC)[27–29,36] is characterized by a coherent superposition of electrons and holes in K and K' valleys. Due to the spin-valley lock, it can also be understood as an easy plane magnet in spin-valley space. The analogy between an easy plane magnet and spin "superfluid" has been discussed in various literatures[42,64,65]; and its extension to IVC state as a spin-valley superfluid is detailed in Ref. [36]. Briefly speaking, an IVC state is described by an order parameter

$$|\Delta|e^{i\theta} \propto \langle c_{\mathrm{K}}^{\dagger} c_{\mathrm{K'}} \rangle \tag{4}$$

which is analogous to the order parameter of a BCS superconductor after particle-hole transformation. The "Cooper pair" here is an intervalley exciton, which carries net spin but not net charge. Therefore, the IVC state can be understood as a spin analogy of superconductor, where $S_z$ takes the role of charge density.

Alternatively, the analogy between IVC state and superfluid can also be seen from semiclassical fluid mechanics of superfluid. The Hamiltonian of an ideal easy-plane magnet can be generally written as[64]:

$$H = \int d^3\boldsymbol{r}[A(\boldsymbol{\nabla}\theta)^2 + K(S^z)^2]/2 \tag{5}$$

where $\theta$ describes the in-plane angle and $S^z$ is the out-of-plane component. The first and second terms describe exchange interaction and magnetic anisotropy, respectively. Eq. 5 takes the same form as that describing superfluid[42], indicating their close analogy.

To allow direct simulation of experimental transport, we construct a minimum hydrodynamic theory for collective modes in the IVC state. The slow variables are the complex IVC order parameter $\Delta(\mathbf{r},t) = |\Delta|e^{i\theta}$and the out-of-plane spin-valley polarization $S^z(\mathbf{r},t)$. We define fluctuations around the equilibrium state at given $(B_z, T)$,

$$\delta S^z(\mathbf{r},t) = S^z - S_0^z, \delta(\mathbf{r},t) = |\Delta| - \Delta_0 \tag{6}$$

and treat the superfluid stiffness $\rho_s$ as a constant over the parameter window of interest, neglecting the normal sector so that spin–valley transport is entirely carried by the condensate.

In the static, spatially uniform limit the free-energy density is a function $f(S^z, |\Delta|, B_z)$. We do not need its microscopic form, only its quadratic expansion around $(S_0^z, \Delta_0)$. In the absence of a displacement field, our system enjoys a $C_{2y}$ symmetry, under which $S^z \to -S^z, B_z \to -B_z, |\Delta| \to |\Delta|$. This symmetry constrains the lowest-order coupling between $\delta S^z$, $\delta$ and $B_z$. Expanding $f$ to second order in $(\delta S^z, \delta)$, we write

$$f = \frac{(\delta S^z)^2}{2\chi_s} + \frac{M_\Delta^2}{2}\,\delta^2 - G(B_z)\,\delta S^z \delta,$$

where $\chi_s$ is the spin-valley compressibility of the superfluid sector, $M_\Delta$ is the curvature associated with the amplitude ("Higgs-like") mode, and $G(B_z)$ is the static mixing between $\delta S^z$ and $\delta$. The above symmetry enforces $G(B_z) = -G(-B_z)$, so for small fields we parameterize $G(B_z) \simeq g\,B_z$ with a single phenomenological coefficient $g$. The full Lagrangian of the system then reads:

$$\mathcal{L} = \left[\hbar(\delta S^z)\dot{\theta} - \frac{(\delta S^z)^2}{2\chi_s} - \frac{1}{2}\rho_s(\nabla\theta)^2\right] + \left[\frac{1}{2}\chi_\Delta(\dot{\delta})^2 - \frac{1}{2}\chi_\Delta v_\Delta^2(\nabla\delta)^2 - \frac{M_\Delta^2\delta^2}{2}\right] + G(B_z)\delta(\delta S^z) \quad (7)$$

where $\theta(\mathbf{r},t)$, $\delta S^z(\mathbf{r},t)$ and $\delta(\mathbf{r},t)$ are promoted to time-dependent fields to describe dynamics. The Berry-phase term $\delta S^z\dot{\theta}$ encodes the canonical structure between the condensate phase and the spin-valley polarization, while the amplitude mode acquires a kinetic term with inertia $\chi_\Delta$. By adding phenomenological damping and a source term that models the effect of the pump on the gap, we arrive at the linearized equations of motion

$$\begin{cases} \hbar\,\partial_t\delta S^z - \nabla\cdot(\rho_s\nabla\theta) = -\hbar\Gamma_S\delta S^z, \\ \chi_\Delta(\partial_t^2\delta + \Gamma_\Delta\,\partial_t\delta - v_\Delta^2\nabla^2\delta) + M_\Delta^2\delta = G(B_z)\,\delta S^z + Q_\delta(\mathbf{r},t), \\ \hbar\,\partial_t\theta = \dfrac{\delta S^z}{\chi_s} - G(B_z)\,\delta. \end{cases}$$

The first equation is the continuity equation for the superfluid spin-valley polarization with current $\boldsymbol{J}_s = -\rho_s\nabla\theta$, supplemented by a local relaxation term with rate $\Gamma_S$. The second describes a damped Higgs-like amplitude mode with damping rate $\Gamma_\Delta$, characteristic velocity $v_\Delta$ and curvature $M_\Delta$, driven by its coupling to $\delta S^z$ and by the pump term $Q_\delta(\mathbf{r},t)$. The last equation is a Josephson-like relation between the phase velocity and the local spin-valley polarization $\delta S^z/\chi_s$, modified by the amplitude-pseudospin mixing. The velocity of the phase mode $v_G$ satisfies $v_G = \sqrt{\rho_s/(\hbar^2\chi_s)}$. In the simulations, we model the effect of pump by a local suppression in the IVC order parameter through local heating and take $Q_\delta(\mathbf{r},t)$ to be a Gaussian in space and time determined by the beam spot.

Extended Data Fig. 7b shows simulated space-time evolution of $\delta S^z$. The simulation parameters are chosen to match the experimental propagation speed; and non-determinable parameters are set to order 1 (see figure caption for a complete list of simulation parameters). While additional ingredients such as a normal sector will further dress the real dynamics, this minimal two field hydrodynamic model already captures the key features of the data: one fast, Goldstone like mode and one slow, Higgs like mode with opposite spin-valley signatures.

**Magnetic field dependence of ordinary and exotic modes:** To further confirm nature of the observed exotic modes, we measured the spatiotemporal transport of the exotic modes (Extended Data Fig. 4a) and the ordinary mode (Extended Data Fig. 4b) as a function of $B_z$. From these measurements, we extracted the field-dependent amplitude of both fast and slow exotic modes (Fig. 5 and Extended Data Fig. 4c). This directly excludes the origin of exotic mode signals from paramagnetic responses and large magnetic susceptibility at the VHS. In that case, their magnetic field dependence should directly follow the system's equilibrium spin-valley polarization and therefore MCD. Experimentally, however, they show distinct behaviors (Fig. 5): the equilibrium MCD increases monotonically and starts to saturate at >2 T. In contrast, the amplitudes of both the fast and slow exotic modes exhibit non-monotonic behaviors: they completely saturate at small magnetic field (1 to 2 T) and are then suppressed at large magnetic field. These observations exclude exotic modes as single particle excitations polarized by the field.

Next, we compare the magnetic field dependence between the ordinary and exotic modes, which reveal striking differences (Extended Data Fig. 4c): the ordinary mode changes monotonically with the magnetic field, and opposite pump helicities (LCP vs. RCP) give opposite trends; while the exotic modes show non-monotonic magnetic dependence. Such contrasting behaviors confirm their distinct nature and provide strong and independent evidence that the exotic modes are collective excitations associated with IVC states. The behavior of ordinary mode is well-understood from its optical injection mechanism[39,41]: The optical selection rule of TMD allows selective injection of electron-hole pairs (excitons) in one valley (e.g. K valley) using circularly polarized (e.g. RCP) light. At zero magnetic field, electrons recombine with holes in both K and K′ valleys with nearly equal probability in a hole doped system because both valleys have a large amount of electrostatically doped holes. After electron-hole recombination, the system is left with pure spin-valley imbalance with zero net charge, which corresponds to the "ordinary" single-particle mode. A magnetic field $B_z$ partially polarizes the hole population into one valley. Depending on the sign of $B_z$, photo-excited electrons have a higher probability of recombining with holes in the K (K') valley owing to its larger hole population, resulting in a smaller (larger) valley imbalance. In the extreme case of complete valley polarization, an initial exciton in the K valley would lead to net valley imbalance of zero (two) holes after electron-hole recombination, as the electron can only recombine with a hole in the K (K') valley. This matches very well with our experiment (Extended Data Fig. 4c).

In contrast, the nontrivial magnetic field dependence of exotic modes provides key evidence of an IVC state. The initial increase at low magnetic field is generally expected, as unpolarized pump light does not break $S_z \leftrightarrow -S_z$ symmetry, which, together with a $S_z$-symmetric ground state, prevents the generation of any spin-valley signal without $B_z$. This can also be understood from that the $S_z$ signal from any excitations is cancelled by their time-reversal pairs, and a finite $B_z$ breaks such symmetry to make them detectable. On the other hand, the suppressing of signal at $B_z$ >1.5T indicates that the exotic modes themselves are suppressed by $B_z$ and eventually disappear when the equilibrium MCD starts to saturate and the system approaches a valley polarized state. This is exactly what we expect from an IVC state: a small $B_z$ would tilt the spin-valley direction to slightly out-of-plane while maintaining the continuous in-plane rotation symmetry; the collective modes themselves are only weakly affected and the observed net $S_z$ signal increases linearly with $B_z$. Upon further increasing $B_z$, the collective modes start to get suppressed as the tilting becomes non-perturbative; meanwhile the observed $S_z$ signal increases sub-linearly before turning around. Eventually, the IVC order is suppressed as the system transitions into valley polarized state, so do the associated collective modes. Other common spin-valley states, such as paramagnetic states and FM-*z* order, are incompatible with our observations. Our magnetic field dependent measurements therefore provide strong evidence of an IVC ground state.

**Possible types of IVC states:** Our observations indicate the emergence of a Goldstone mode in the spin-valley space, which is a generic feature of IVC orders. Therefore, our results cannot distinguish between different types of IVC states at a qualitative level. In particular, an IVC state (superposition between states different by momentum **K**-**K**') and an IKS state (superposition between states different by momentum **K**-**K**'+**q**) both spontaneously break the continuous U(1) valley rotation symmetry and give rise to a gapless Goldstone mode. Therefore, their difference is primarily of quantitative instead of qualitative nature. In twisted graphene[34,35], the most energetically favorable **q** takes a value of roughly 1/3 of moire Brillouin zone. This leads to considerable modulation of charge textures, such as breaking of rotation symmetry. In twisted TMD, on the other hand, theoretical investigation of IKS states is still at an early stage, and it remains unclear when and which IKS states will be favored. Quantitative comparison between our measurements and theoretical modeling may allow distinguishing between IVC and IKS states and is an interesting future direction.

**Generality of our ultrafast spin imaging technique:** Our approach can be readily applied to other moiré TMD systems. For example, our approach provides a powerful tool to investigate collective excitations associated with fractional Chern insulator (FCI) states in twisted $MoTe_2$. FCI states are known to host exotic collective excitations, such as fractionalized excitations[66] and magnetorotons[67]. Directly probing and isolating these excitations have been challenging, because many of them are charge-neutral and therefore show weak responses in electrical measurements. Furthermore, they typically coexist with other conventional excitations that show stronger responses, such as charged quasiparticles, and therefore are eclipsed by the latter in steady-state measurements that measure total responses. Our approach can potentially overcome both challenges. We probe collective excitations by encoding them into nearby excitons and reading out optically. Such "exciton sensing" concept is quite general as there exist many interaction channels between collective excitations and excitons, such as dielectric screening in graphene[68,69] and phase space filling in twisted TMD here. In addition, the non-equilibrium transport measurement performed in the present work allows us to directly capture and separate multiple coexisting modes from their distinct propagation, such as the slow and fast exotic modes observed here. We emphasize that this is only possible with both space and time information, and therefore cannot be achieved with only the temporal information (such as conventional pump-probe spectroscopies) or only the spatial information (such as scanning probe measurements).

## Extended Data Figure 1

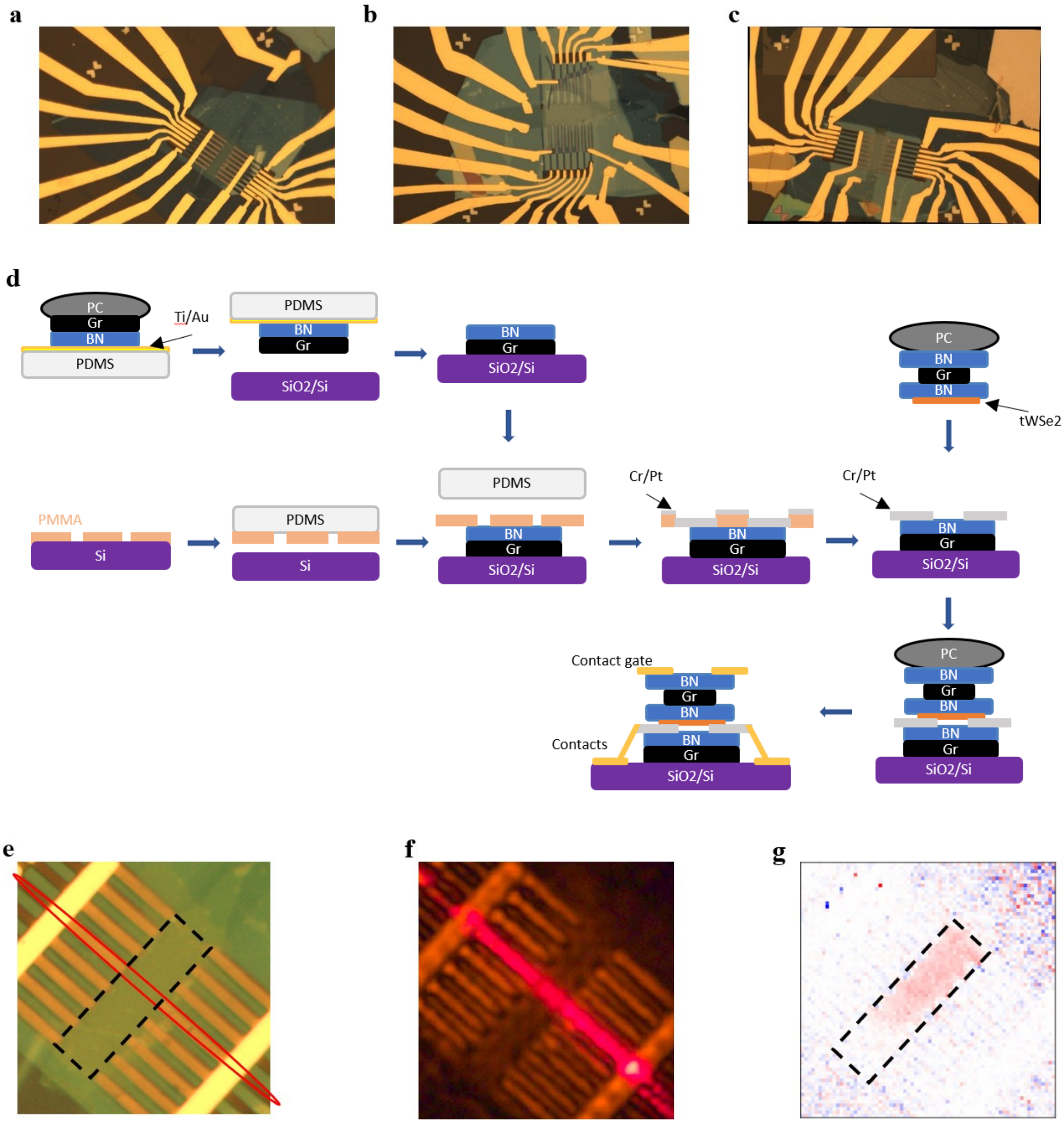

**Extended Data Figure 1. Sample preparation and measurement configurations. a-c,** Optical microscope images of device D1-D3. **d,** Sample fabrication step-by-step diagram. Upper panel: Gold coated PDMS assisted flipping of bottom gate stack. Middle panel: a PMMA stencil mask is used for metal deposition so that a polymer-free Pt surface is achieved. Lower panel: $tWSe_2$ stack is released to the bottom gate. A 5 nm Pd layer is deposited on top as local contact gate. **e,** Optical microscope images of device D1 under white light illumination. The black dashed contour outlines the twisted $WSe_2$ region. The red solid ellipse denotes shape of the pump. **f,** Device image under actual measurement configuration illuminated by both pump (line shape, red) and probe (wide-field, orange) lights. **g,** A representative snapshot of the pump probe results.

## Extended Data Figure 2

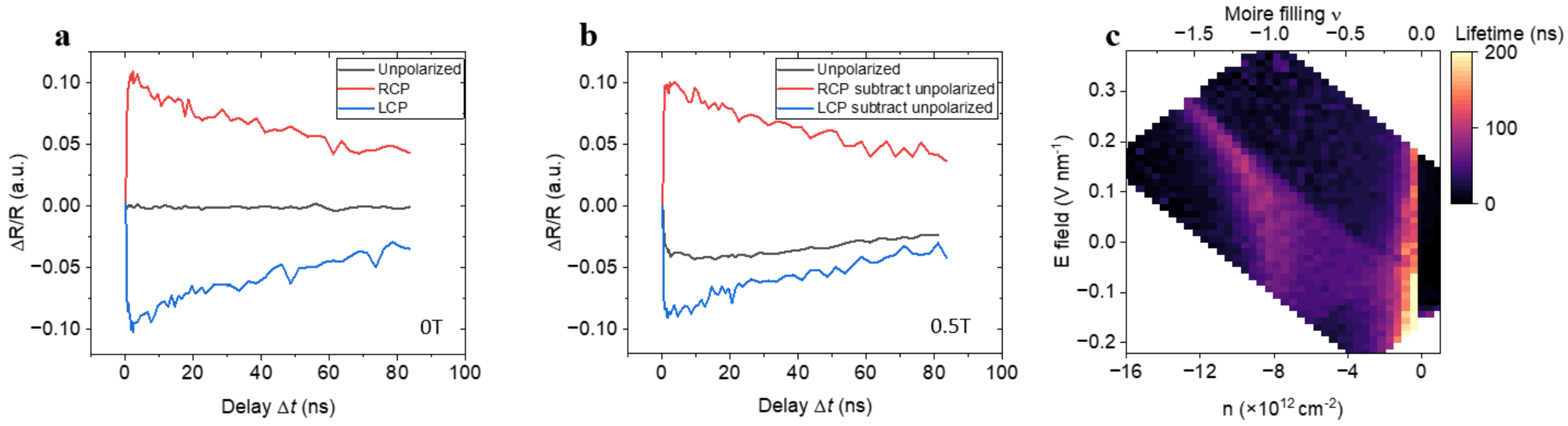

**Extended Data Figure 2. Isolating ordinary and exotic modes. a,** Delay dependent $\Delta R/R$ at $n = -8.3\times10^{12}$ cm$^{-2}$, $E = 0.1$ V nm$^{-1}$ and $B_z = 0$ T, using unpolarized, RCP and LCP pump. No signal is observed under unpolarized pump, while the RCP and LCP pumps give opposite signals. Unpolarized pump is equivalent to linearly polarized pump due to short valley coherence time and non-resonant excitation. **b,** Same as **a** but under $B_z = 0.5$ T. The ordinary mode is obtained from the signal difference between circular and unpolarized pumps, which remain largely unchanged compared to 0 T. In contrast, unpolarized pump now gives finite signal, corresponding to the exotic mode. **c,** Spin-valley lifetime of ordinary modes across the ($n$, $E$) phase space at $B_z = 0$ T, which is almost identical to 0.5 T (Fig. 1g).

## Extended Data Figure 3

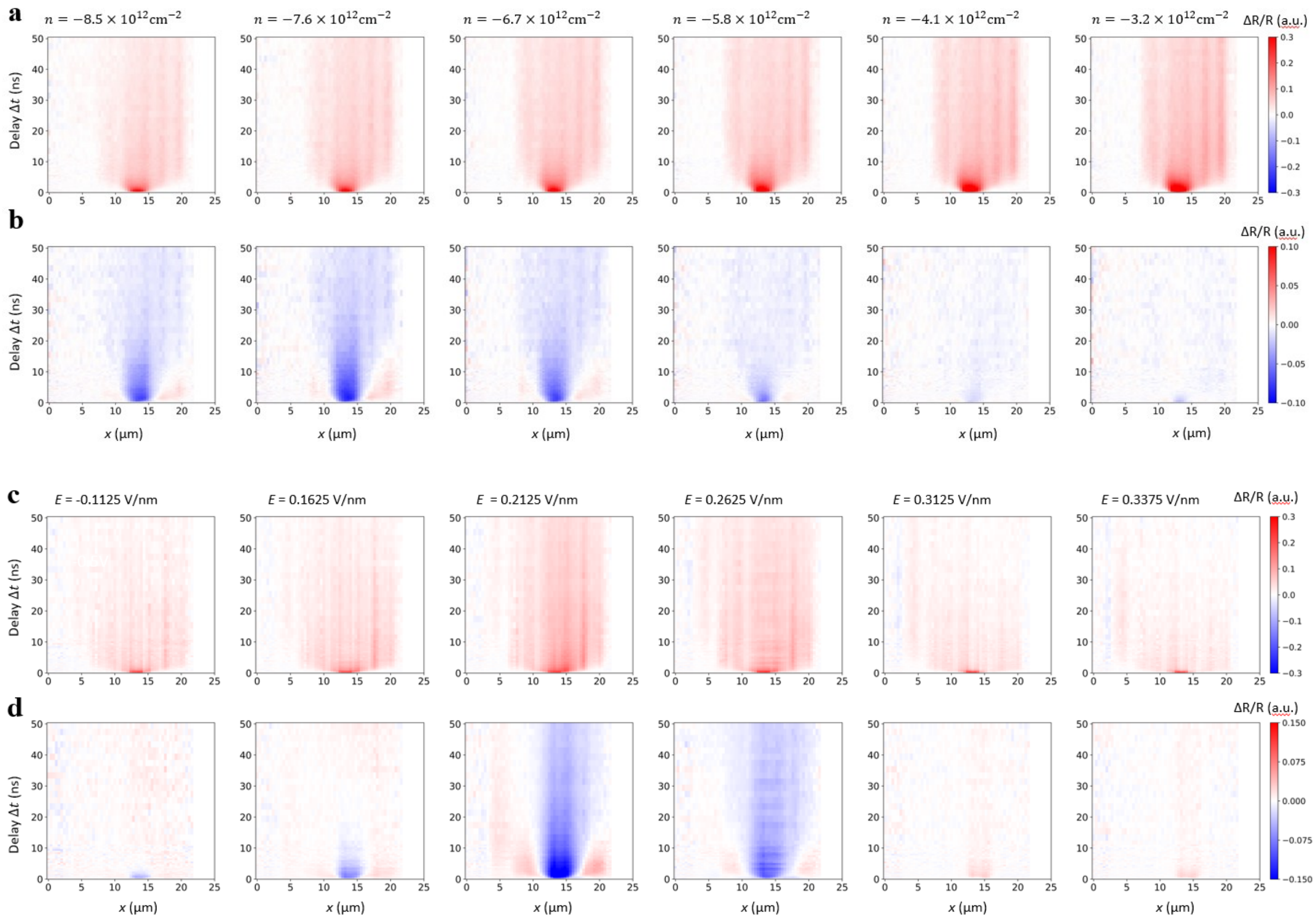

**Extended Data Figure 3. Additional spin-valley transport results from device D1. a,b,** Delay-dependent spatial profiles at $E = -0.025$ V/nm, $B_z = 0.5$ T and representative doping for the ordinary (**a**) and exotic (**b**) modes. **c,d,** Delay-dependent spatial profiles at $n = -10.2\times10^{12}$ cm$^{-2}$, $B_z = 0.5$ T and representative electric field for the ordinary (**c**) and exotic (**d**) modes. Exotic modes disappear outside the VHS region, while the ordinary mode is observed at all doping/electric fields.

**Extended Data Figure 4**

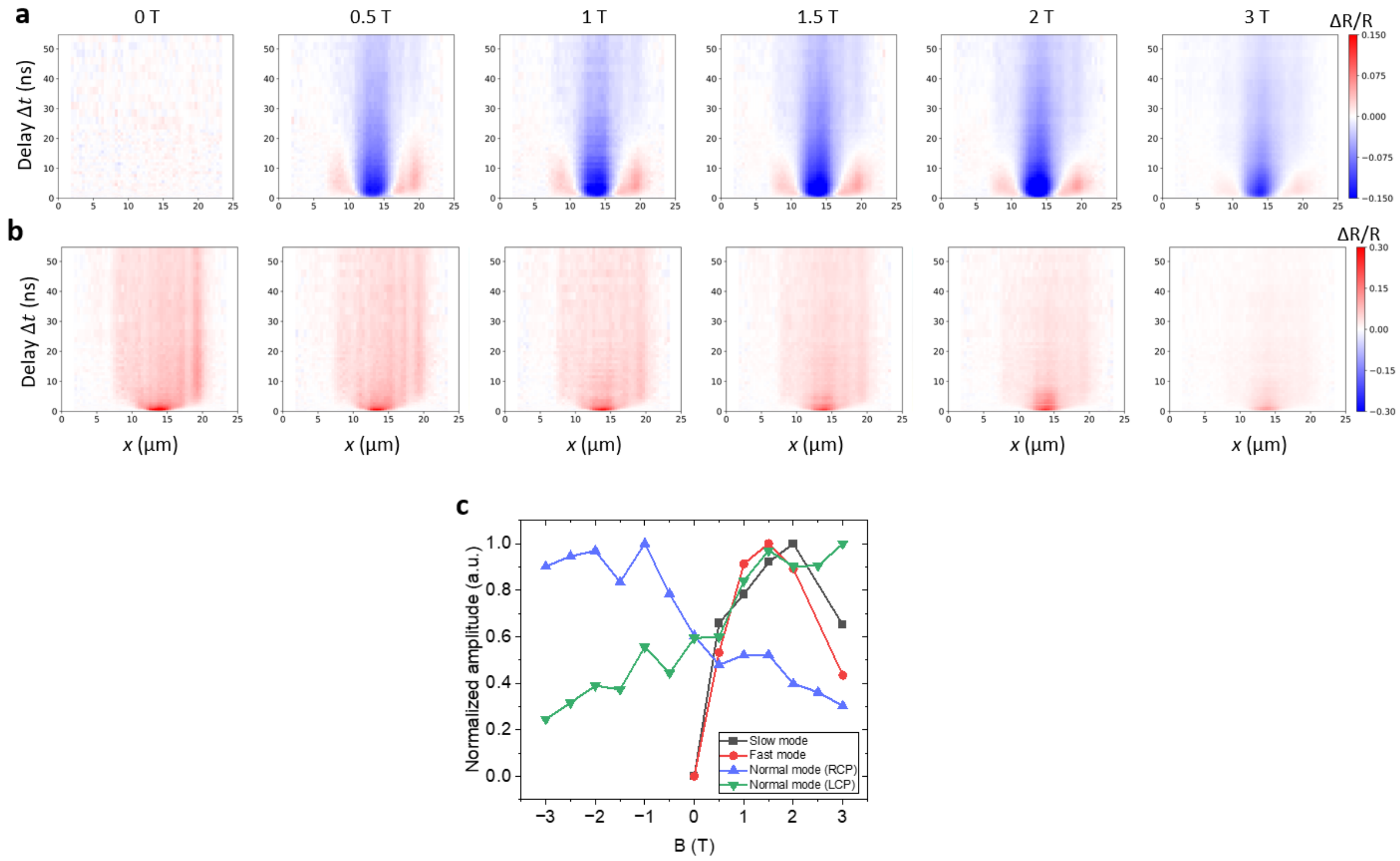

**Extended Data Figure 4. Magnetic field dependence of spin-valley transport from device D1. a,b,** Delay-dependent spatial profiles of exotic modes (**a**) and ordinary mode (**b**) at $n = -8.3\times10^{12}$ cm$^{-2}$, $E = 0.1$ V nm$^{-1}$ and different magnetic fields from 0 to 3 T. **c,** Magnetic field dependent amplitudes of the slow and fast exotic modes comparing with that of the normal mode excited by RCP and LCP pump. The amplitude of normal modes is extracted from wide-field pump configuration.

**Extended Data Figure 5**

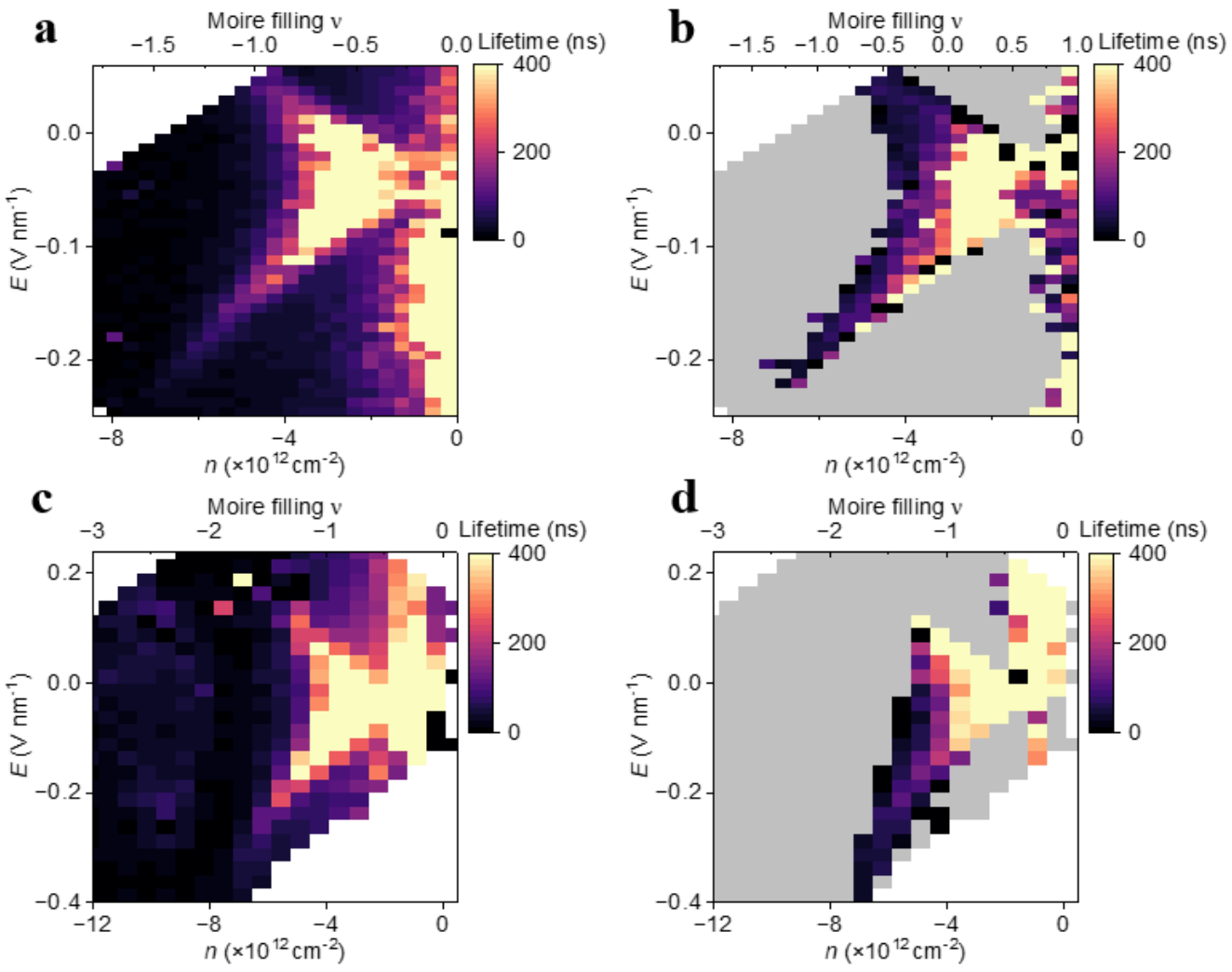

**Extended Data Figure 5. Additional results from devices D2 (3.8°) and D3 (3.5°). a,b,** Spin-valley lifetime of ordinary (**a**) and exotic (**b**) modes across the ($n$, $E$) phase space at $B_z$ = 0.5 T for device D2, both showing enhanced lifetime around VHS. **c,d**, Same as (**a**) and (**b**) for device D3.

## Extended Data Figure 6

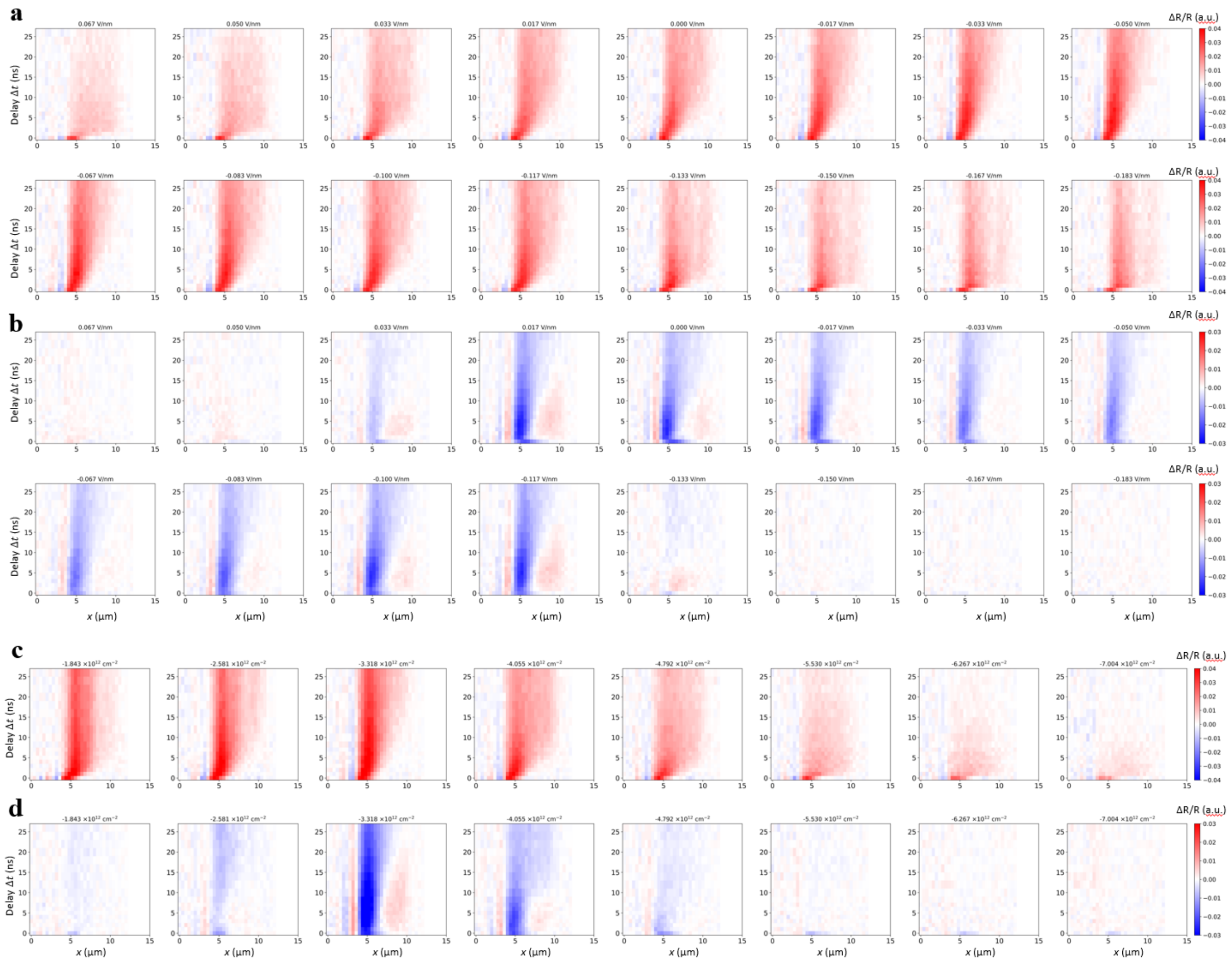

**Extended Data Figure 6. Additional spin-valley transport results from device D2. a,b,** Delay-dependent spatial profiles at $n$ = -3.7×10$^{12}$ cm$^{-2}$, $B_z$ = 0.5 T and representative electric field for the ordinary (**a**) and exotic (**b**) modes. **c,d,** Delay-dependent spatial profiles at $E$ = −0.1 V/nm, $B_z$ = 0.5 T and representative doping for the ordinary (**c**) and exotic (**d**) modes. Exotic modes disappear outside the VHS region, while the ordinary mode is observed at all doping/electric fields.

**Extended Data Figure 7**

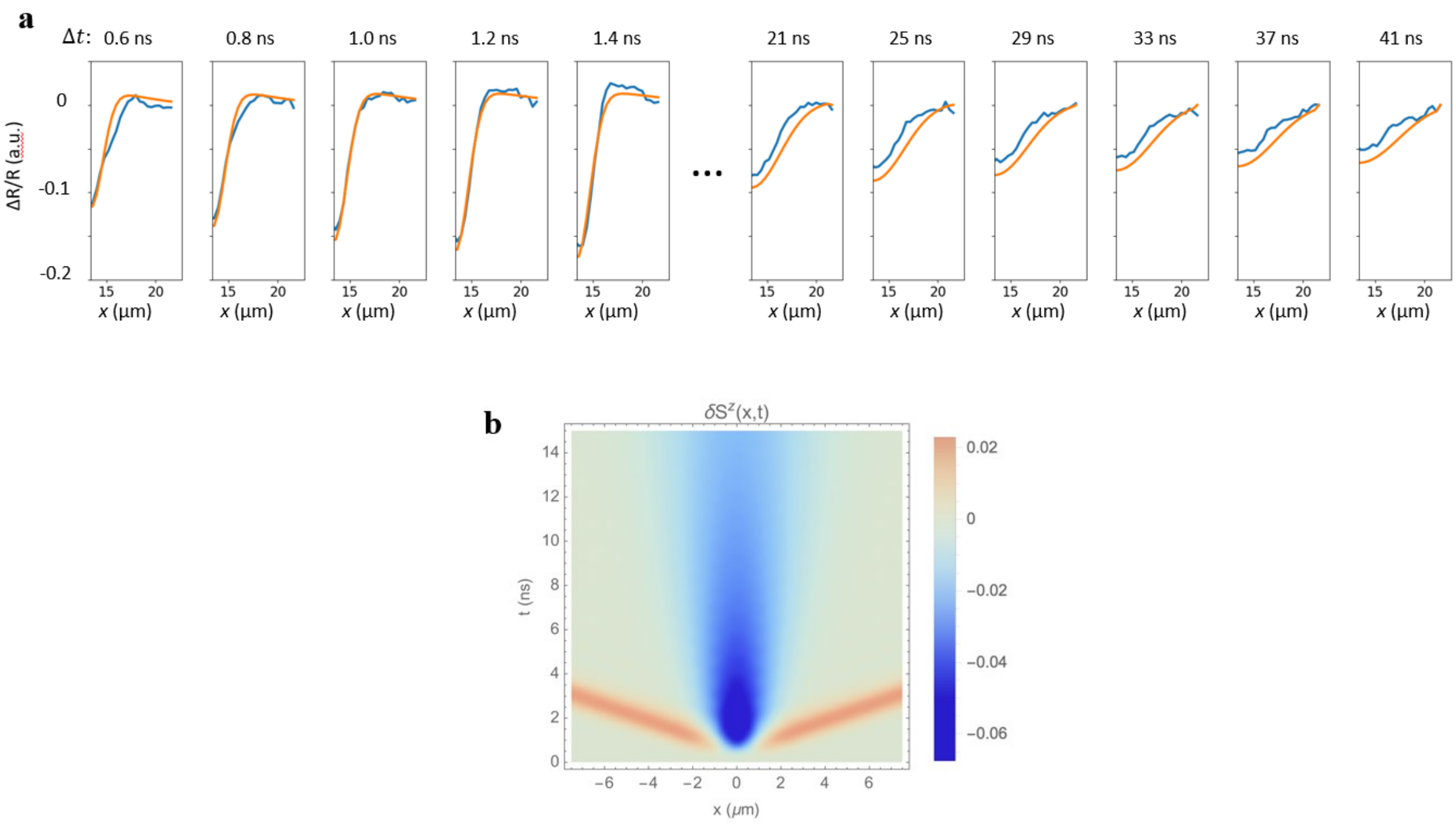

**Extended Data Figure 7. Diffusive fitting of exotic modes and minimum hydrodynamic theory. a,** Two-component diffusive-decay fitting of the exotic modes using data within the first two nanoseconds. With a diffusion constant of 200 $cm^2/s$, the fitting (orange) captures initial evolution of the fast mode better but still cannot fully describe the experimental spatial-temporal profile (blue). **b,** Simulated transport of exotic modes from minimum hydrodynamic theory. For the Goldstone mode sector, the superfluid stiffness $\rho_s \approx 2k_B T_c/\pi \approx 0.5$ meV ($T_c \approx 10$ K). The spin compressibility $\hbar\chi_s = \rho_s/(\hbar v_G^2) \approx 84$ ns/$\mu m^2$, with $v_G \approx 3$ μm/ns for the fast exotic mode. Spin polarization damping rate $\Gamma_\Delta \approx 0.01$ $ns^{-1}$, estimated from experimental spin lifetime. For the amplitude mode sector, the effective diffusion constant $D_\Delta \approx v_\Delta^2/\Gamma_\Delta$. Taking the velocity $v_\Delta$ to be similar of $v_G$ and using the experimentally extracted diffusion constant $D_\Delta \approx 2$ $cm^2/s$, the damping rate $\Gamma_\Delta = v_\Delta^2/D_\Delta \approx 45$ $ns^{-1}$. The inertia $\chi_\Delta$ and amplitude curvature $M_\Delta^2$ cannot be determined from experiment and are scaled to be order 1. The pump excitation $Q_\delta(\mathbf{r}, t)$ takes a Gaussian in space and time of $e^{-x^2/(2\sigma_X{}^2)} * e^{-t^2/(2\sigma_T{}^2)}$, where $\sigma_X = 0.6$ μm and $\sigma_T = 0.5$ ns. Phenomenological coefficient $g$ is taken as 0.05. The model assumes propagation to infinity and neglects boundary effects.

**Extended Data Figure 8**

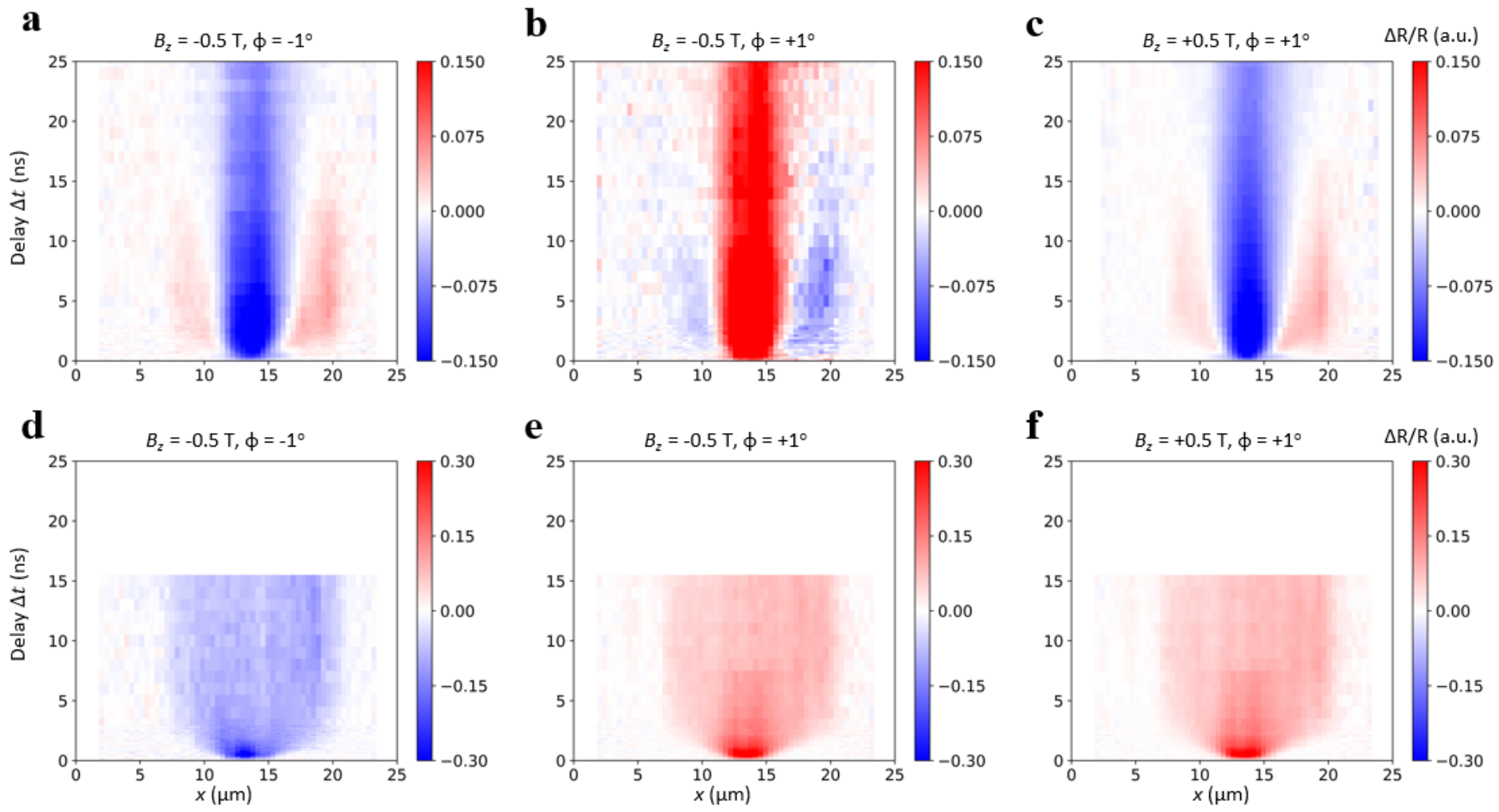

**Extended Data Figure 8. Selective probe of spin-valley excitations. a-c,** Delay-dependent spatial profiles of exotic modes at magnetic field $B_z$ = -0.5 T and detection angle $\phi$ = -1° (**a**), $B_z$ = -0.5 T, $\phi$ = +1°(**b**), $B_z$ = +0.5 T, $\phi$ = +1° (**c**). **d-f**, Same measurements for ordinary mode. Both ordinary and exotic signals change sign with $\phi$, confirming that all are spin-valley excitations (see Methods). Meanwhile, only the exotic modes change sign upon reversing the magnetic field while the ordinary mode does not, indicating their distinct nature. For all measurements in the main text, we use $\phi$ = +1°.

## Extended Data Figure 9

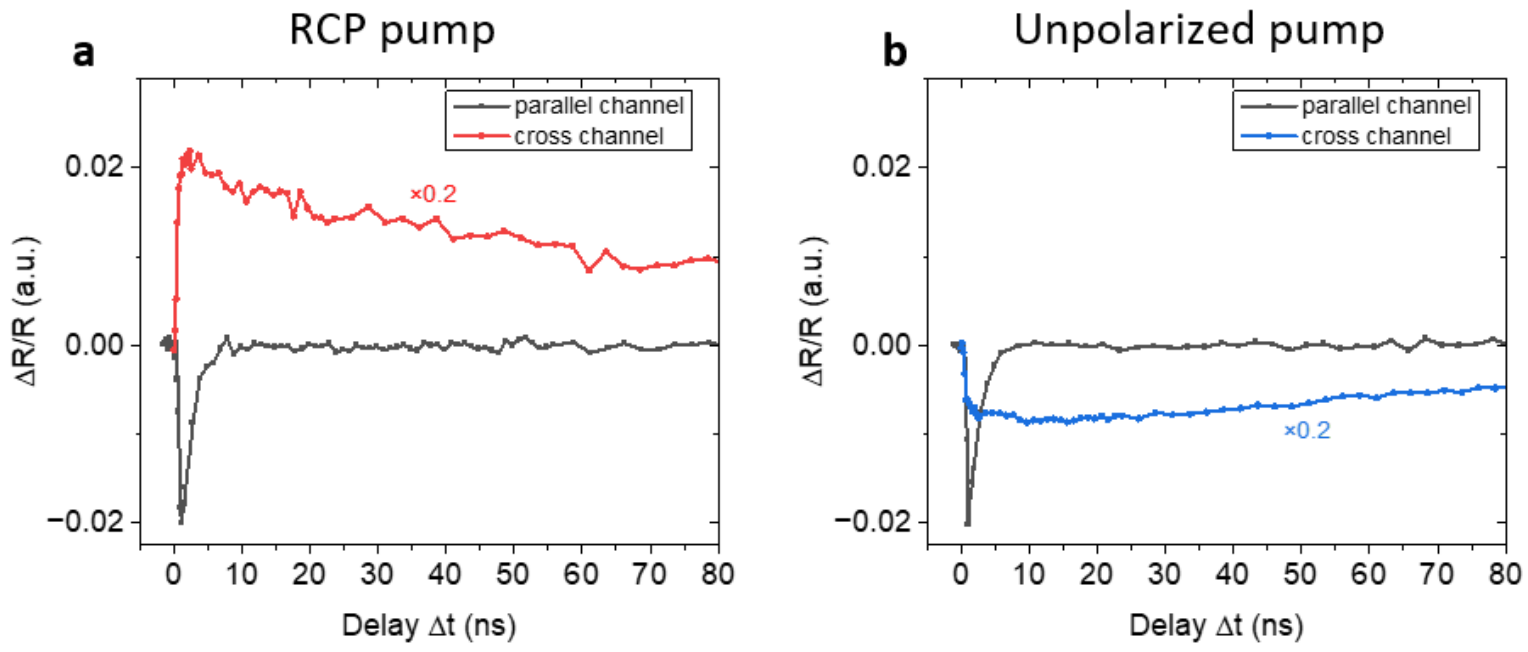

**Extended Data Figure 9. Isolating charge and spin-valley responses. a**, Delay-dependent ΔR/R at $n = -8.3\times10^{12}$ cm$^{-2}$, $E = 0.1$ V nm$^{-1}$ and $B_z = 0$ T under wide-field RCP pump using the cross and parallel detection channel. The "parallel" configuration detects the total reflection of two valleys and captures charge (population) dynamics, while the cross channel detects the valley-contrasting signals and therefore spin-valley excitations (see Methods). **b**, Similar as in (a) with unpolarized pump and at $B_z = 0.5$ T. The lifetimes of both normal mode (**a**) and exotic modes (**b**) are much longer than that of charge (population) excitation, indicating that they are pure spin-valley excitations decoupled from charge excitations. The signals in the cross channel are scaled by 0.2 for clarity.

**Extended Data Figure 10**

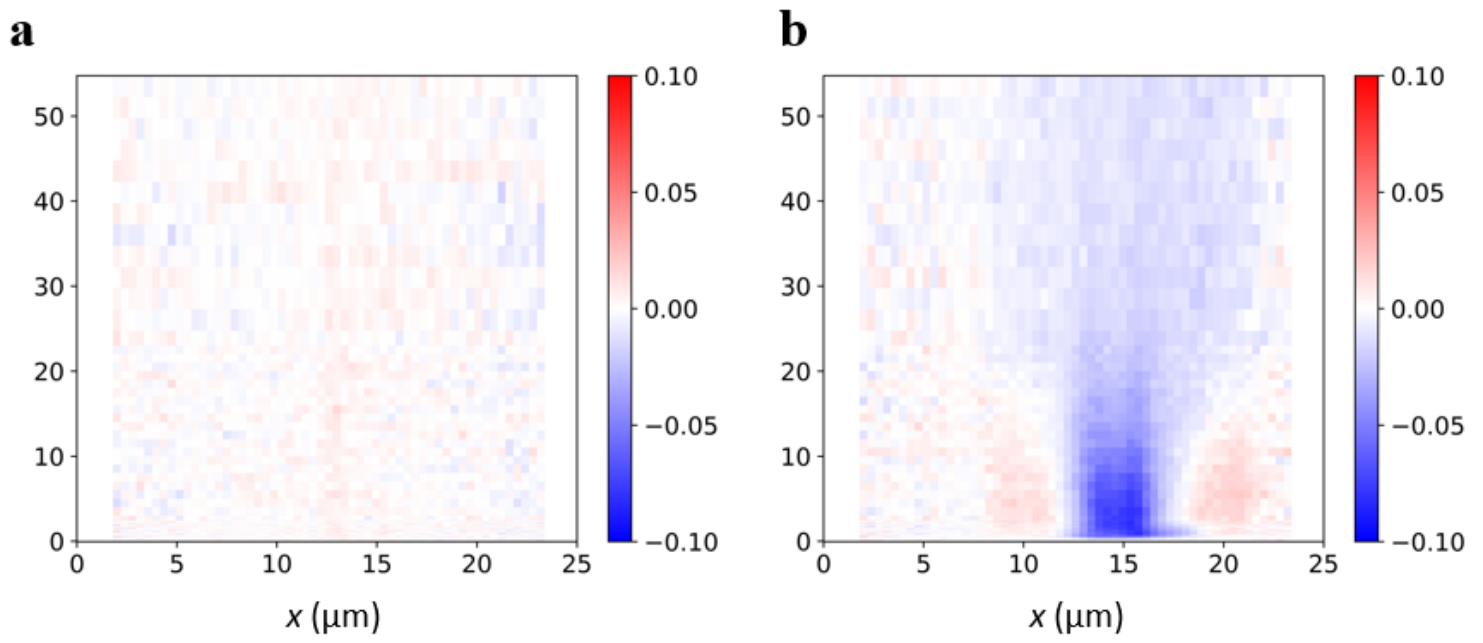

**Extended Data Figure 10. Measurements with below-gap pump. a,b,** Delay-dependent spatial profiles at $n = -10.2\times10^{12}$ cm$^{-2}$, $E = 0.26$ V nm$^{-1}$ and $B_z = 0.5$ T for the ordinary (**a**) and exotic (**b**) modes using a below-gap (900 nm) pump. The ordinary mode cannot be excited anymore, whereas exotic modes are still observed. This confirms that the exotic modes are low-energy collective modes instead of non-collective quasiparticles from interband particle-hole excitations.